\documentclass[a4paper,11pt]{article}

\usepackage{jheppub} 
\usepackage{lineno}
\usepackage{cancel}
\usepackage{graphicx}
\usepackage{dcolumn}
\usepackage{bm}
\usepackage[usenames,dvipsnames]{xcolor}
\usepackage{soul}
\usepackage{enumitem}
\usepackage{xspace}
\usepackage{braket}
\usepackage{booktabs}
\usepackage{lipsum}
\usepackage[normalem]{ulem}
\usepackage[ISO]{diffcoeff}
\usepackage{subcaption}
\usepackage{ragged2e}
\usepackage{mathrsfs}
\usepackage{amsmath}
\usepackage{amssymb}
\usepackage{mathbbol} 
\usepackage[titletoc]{appendix}
\usepackage{esvect} 
\usepackage{titlesec}
\usepackage{titletoc}
\usepackage{chngcntr}
\usepackage{fontawesome5}
\usepackage{mleftright}
\usepackage[nodayofweek]{datetime}
\usepackage{comment}
\definecolor{darkblue}{rgb}{0,0,0.5}
\definecolor{darkgreen}{rgb}{0.0,0.5,0.2}
\definecolor{darkred}{rgb}{0.6,0,0}
\usepackage{hyperref}
\usepackage[capitalise]{cleveref}
\usepackage{orcidlink}
\usepackage{gensymb} 
\usepackage{float}

\usepackage{tikz}
\usetikzlibrary{calc,angles,quotes,intersections,arrows.meta,decorations.pathreplacing}

\Crefname{appsec}{Appendix}{Appendices}

\usepackage{listings}
\usepackage{xcolor}

\definecolor{vscodeblue}{RGB}{0,0,255}
\definecolor{vscodegreen}{RGB}{0,128,0}
\definecolor{vscodered}{RGB}{163,21,21}
\definecolor{vscodegray}{RGB}{100,100,100}
\definecolor{vscodebg}{RGB}{250,250,250}

\lstdefinestyle{vscode-python}{
    language=Python,
    basicstyle=\ttfamily\footnotesize,
    keywordstyle=\color{magenta},
    commentstyle=\color{vscodegreen},
    stringstyle=\color{vscodered},
    numberstyle=\color{blue},
    showstringspaces=false,
    breaklines=true,
    keepspaces=true,
    frame=lines,
    rulecolor=\color{black!60},
    xleftmargin=0.5em,
    framexleftmargin=0.5em,
    tabsize=4,
    columns=fullflexible,
    upquote=true
}

\graphicspath{{figures/}}

\definecolor{firebrick}{HTML}{B22222}
\definecolor{orcid-green}{RGB} {166, 206, 57}
\hypersetup{
    colorlinks=true,       
    linkcolor=firebrick,          
    citecolor=firebrick,        
    filecolor=firebrick,      
    urlcolor=firebrick          
}

\newcommand{\dd}{\mathop{}\!\mathrm{d}}
\newcommand{\mrm}[1]{\mathrm{#1}}
\newcommand{\usim}{\mathord{\sim}}

\newcommand{\snudd}{\texttt{SNuDD}\xspace}
\newcommand{\dCP}{\delta_\mathrm{CP}}
\newcommand{\cevns}{CE$\nu$NS\xspace}
\newcommand{\eves}{E$\nu$ES\xspace}
\newcommand{\nsi}[1]{\varepsilon^{\eta,\varphi}_{#1}\xspace}
\newcommand{\tnyr}{{\rm tonne\text{-}years}\xspace}

\title{\boldmath \snudd: Solar Neutrinos for Direct Detection}

\preprint{IFT-UAM/CSIC-26-97}

\author[a]{Dorian W.~P.~Amaral\,\orcidlink{0000-0002-1414-932X},}
\emailAdd{damaral@ifae.es}

\author[b]{David Cerde\~no,\,\orcidlink{0000-0002-7649-1956}}
\emailAdd{davidg.cerdeno@ift.csic.es}

\author[c,d]{Andrew Cheek\,\orcidlink{0000-0002-8773-831X},}
\emailAdd{acheek@sjtu.edu.cn}

\author[b,e]{Valeria Costa\,\orcidlink{0009-0000-8129-5741},}
\emailAdd{valeria.costa@ift.csic.es}

\author[b]{and Patrick Foldenauer\,\orcidlink{0000-0003-4334-4228}}
\emailAdd{patrick.foldenauer@cisc.es}

\affiliation[a]{Institut de F\`{ı}sica d’Altes Energies (IFAE), The Barcelona Institute of Science and Technology,
Campus UAB, 08193 Bellaterra (Barcelona), Spain}

\affiliation[b]{Instituto de F\'isica Te\'orica IFT-UAM/CSIC, 
Cantoblanco, E-28049, Madrid, Spain}

\affiliation[c]{Tsung-Dao Lee Institute \& School of Physics and Astronomy, Shanghai Jiao Tong
University, Shanghai 200240, China}

\affiliation[d]{Key Laboratory for Particle Astrophysics and Cosmology (MOE) \& Shanghai Key Laboratory for Particle Physics and Cosmology, Shanghai Jiao Tong University, Shanghai 200240, China}

\affiliation[e]{Departamento de Física Te\'orica, Universidad Aut\'onoma de Madrid, Cantoblanco, E-28049, Madrid, Spain}

\abstract{We introduce Solar Neutrinos for Direct Detection (\snudd~\href{https://github.com/SNuDD/SNuDD}
{\faGithub}): an open-source Python package that enables the computation of the solar neutrino rate spectrum at direct detection experiments. \snudd can be used to determine
the differential rate for both nuclear and electron recoils within the Standard Model and in the presence of beyond Standard Model physics effects, such as those arising from neutral-current non-standard interactions (NSI). The package accounts for matter effects during neutrino propagation through both the Sun and the Earth and for modifications to the scattering cross sections at the interaction site. We employ \snudd to place new limits on the effective NSI couplings using results from the xenon-based direct detection experiments LZ, XENONnT, and PandaX-4T, and we project the sensitivity of a future xenon detector based on the planned XLZD and PandaX-xT observatories. We find that current direct detection experiments are rapidly approaching sensitivities comparable to those of dedicated neutrino experiments and that future xenon detectors can provide leading constraints. 
We recommend that \snudd be used to 
combine incoming direct detection data with those from neutrino experiments in future global fits,
placing direct detection within the broader landscape of neutrino physics.
\vspace{1.cm}
\begin{center}
    \includegraphics[width=0.25\linewidth]{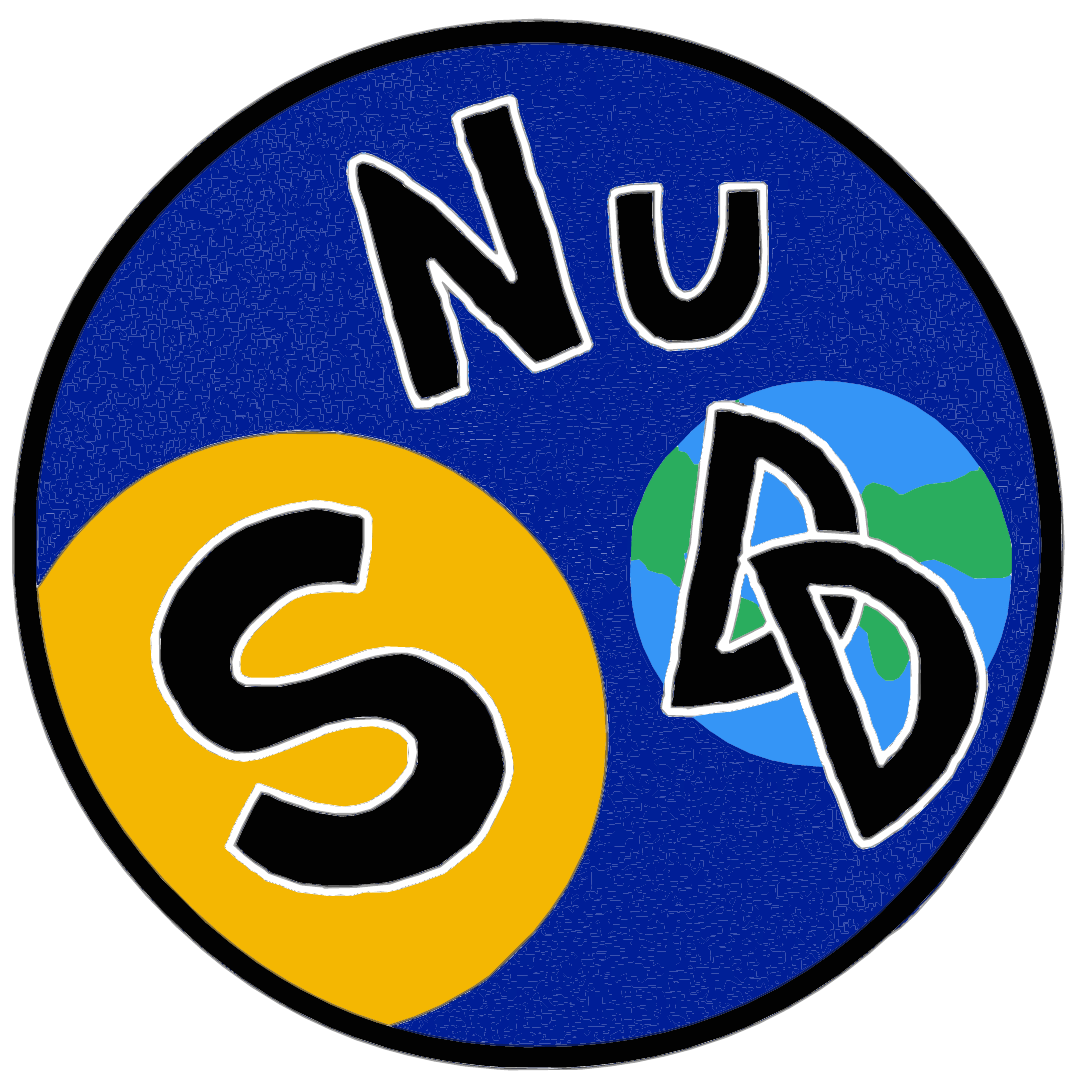}
\end{center}

}

\begin{document}

\maketitle
\flushbottom

\section{Introduction}
\label{sec:intro}

Direct detection experiments have now reached a sensitivity where solar neutrinos, once considered a negligible background, can no longer be ignored. This marks the onset of the neutrino fog, a regime in which interactions from solar neutrinos (or atmospheric neutrinos for higher energies) become increasingly difficult to distinguish from a genuine dark matter signal~\cite{Billard:2013qya, Billard:2014yka, OHare:2021utq, Carew:2023qrj}. However, this background can also be viewed as an opportunity. As the number of detected solar neutrino events increases, direct detection experiments are becoming sensitive neutrino observatories. This opens up exciting possibilities to study Standard Model (SM) properties, explore solar physics, and search for physics beyond the Standard Model (BSM)~\cite{Harnik:2012ni,Baudis:2013qla,Billard:2014yka,Cerdeno:2016sfi,Bertuzzo:2017tuf,Cui:2017ytb,Dutta:2017nht,Newstead:2018muu,Essig:2018tss,AristizabalSierra:2019ykk,Dutta:2019oaj,Amaral:2020tga,Abdullah:2020iiv,Amaral:2021rzw,deGouvea:2021ymm,Schwemberger:2022fjl,Abdullah:2022zue,OHare:2022jnx,Amaral:2023tbs,Amaral:2024edw,Acevedo:2024wmx,DeRomeri:2024dbv,DeRomeri:2024hvc,AtzoriCorona:2025gyz,DeRomeri:2024iaw,DeRomeri:2025nkx,DeRomeri:2026prc,Kelly:2026avh,Arguelles:2026kaz}.

This potential has recently been demonstrated by the first claims of the observation of coherent elastic neutrino–nucleus scattering (\cevns) due to neutrinos from the $^8\mrm{B}$ solar flux by the three world-leading experiments based on xenon time-projection chambers: PandaX-4T~\cite{PandaX:2024muv}, XENONnT~\cite{XENON:2024ijk,XENON:2026ydt}, and LZ~\cite{LZ:2025igz}. In addition, PandaX-4T has measured elastic neutrino-electron scattering (\eves) from $pp$ solar neutrinos, establishing electron recoils as a complementary detection channel~\cite{PandaX:2026qdp}. Together, these results mark the first dedicated measurements of solar neutrinos in dark matter detectors using both nuclear and electron recoils. With next-generation experiments such as XLZD and PandaX-xT on the horizon~\cite{PANDA-X:2024dlo,XLZD:2024nsu}, the role of dark matter detectors as neutrino observatories is expected to continue growing.

Future direct detection experiments that employ other targets may also be able to test solar neutrinos through \cevns or \eves. This is the case, for example, for liquid argon detectors, where experiments such as DEAP-3600 \cite{DEAP-3600:2017ker} and DarkSide-20k \cite{DarkSide-20k:2017zyg} cannot reach the neutrino fog but may join forces with the future Global Argon Dark Matter Collaboration.
Likewise, low-threshold experiments designed to search for low-mass WIMPs, such as SuperCDMS \cite{SuperCDMS:2016wui}, NEWS-G \cite{NEWS-G:2022kon}, DAMIC-M \cite{DAMIC-M:2023gxo}, and RES-NOVA~\cite{Pattavina:2020cqc}, are potentially very interesting since they would be able to probe the solar flux at lower energies, provided that their payload increases sufficiently. 
Moreover, there are also proposals to use directional detectors such as CYGNO to test neutrino BSM physics with electron recoils~\cite{Shekar:2025xhx}.

New physics in the neutrino sector can be described in terms of non-standard interactions (NSI), a model-independent effective framework that preserves the SM vector current structure~\cite{Wolfenstein:1977ue,Guzzo:1991cp,Guzzo:2000kx,Gonzalez-Garcia:1998ryc,Bergmann:2001mt,Guzzo:2001mi,Gonzalez-Garcia:2004pka}. These interactions can modify both neutrino propagation through matter and the neutrino scattering cross section at the detection point. Extracting robust NSI constraints from \cevns and \eves data requires accurate theoretical predictions for the expected recoil spectra, accounting simultaneously for both of these effects. While existing tools such as \texttt{NuSQuIDS}~\cite{Arguelles:2021twb}, \texttt{NuFAST}~\cite{Denton:2024pzc,Denton:2025oxf}, and \texttt{PEANUTS}~\cite{Gonzalo:2023mdh} can be used to compute solar neutrino propagation effects, no public package currently exists that can compute the \cevns and \eves rate spectra in the presence of NSI and account for the full phase-correlation of different solar neutrino flavour states. The recent detection of both solar \cevns and \eves in direct detection experiments highlights the need for a dedicated software package.

In this work, we introduce \snudd, a Python package that computes the solar neutrino recoil spectra at direct detection experiments. In \cref{sec:theory},  we introduce the neutrino NSI formalism and specify the parametrisation used by our code. We then show how NSI impact the propagation of neutrinos through the Sun and the Earth, as well as the scattering cross section at the detection site. In \cref{sec:snudd}, we present \snudd and explain how to use it to determine the neutrino recoil spectrum and detector rates for a given particle model. As a physics application, in \cref{sec:application}, we  employ \snudd to compute new limits on the NSI parameters using the recent \cevns and previous \eves data from LZ, XENONnT, and PandaX-4T, as well as to project a constraint for a future experiment based on XLZD and PandaX-xT.

Our results demonstrate that \snudd can be used to accurately predict new neutrino physics signals at direct detection experiments, readily incorporate new data into BSM neutrino physics landscapes, and easily facilitate the inclusion of these experiments in future global analyses.

\section{Neutrino NSI at Direct Detection Experiments}
\label{sec:theory}

In this work, we study the sensitivity of direct detection experiments to neutrino NSI via solar neutrino scattering with either nuclei, leading to nuclear recoil (NR) events, or electrons, resulting in electron recoil (ER) events. The central quantity that must be evaluated is the differential rate spectrum~\cite{Coloma:2022umy,Amaral:2023tbs}
\begin{align}\label{eq:trace}
    \frac{\dd R}{\dd E_R}(E_R)=n_T \int_{E_\nu^{\min }} \frac{\mathrm{d} \phi_\nu}{\mathrm{d} E_\nu} \operatorname{Tr}\left[\rho \frac{\mathrm{d} \zeta}{\mathrm{~d} E_R}\right] \mathrm{d} E_\nu\,,
\end{align}
which expresses the rate of solar neutrino scattering events occurring per unit recoil energy $E_R$ per unit detector mass. Here, $n_T$ is the  number of scattering targets per unit mass, $\phi_\nu$ is the neutrino flux at the source, $\rho(E_\nu)$ is the neutrino density matrix evolved to the interaction point, and $\dd \zeta / \dd E_R$ denotes the generalised scattering cross section incorporating flavour correlations in the scattering matrix elements between different neutrino flavours. The trace acts as the quantum statistical average of the generalised cross section, $\operatorname{Tr}[\rho\, \zeta]=\langle\zeta\rangle$, capturing the expectation value of the scattering cross section even when the incoming neutrinos are in a mixed state.
The integration is performed over neutrino energies $E_\nu$, running from the minimum neutrino energy needed to induce a recoil of energy $E_R$. Both the number of scattering targets and minimum neutrino energy are target dependent, relying on the mass of the target $m_T$. This is taken to be the mass of the nucleus or the electron depending on whether we are interested in the \cevns or \eves spectrum, respectively, with $E_\nu^\mrm{min}(E_R) = \big(E_R + \sqrt{E_R^2 + 2 m_T E_R}\big)  /2$.

The two key ingredients entering \cref{eq:trace} are the density matrix and generalised cross section. The former captures how the relative neutrino flavour fractions evolve as neutrinos propagate from the Sun, through the Earth, and ultimately to the detector. The latter encapsulates the scattering physics at the point of interaction within the detector. Both of these quantities are modified in the presence of new physics in the neutrino sector.

The NSI framework is a convenient parametrisation of new neutrino physics in terms of a low-energy effective field theory that quantifies modifications of the SM electroweak interactions of neutrinos. It is described by the effective Lagrangian~\cite{Wolfenstein:1977ue,Davidson:2003ha,Gonzalez-Garcia:2004pka, Miranda:2015dra} 
\begin{equation}
	\mathcal{L}_{\rm NSI} = - 2 \sqrt{2}\, G_F \sum_{P, f, \alpha, \beta} \varepsilon^{fP}_{\alpha\beta}\, \left(\bar{\nu}_\alpha \gamma_\mu P_L \nu_\beta\right) \left(\bar{f} \gamma^\mu P f\right)\,,
\end{equation}
where $G_F$ is the Fermi constant and $\varepsilon^{fP}_{\alpha\beta}$ are the NSI coupling coefficients. The sum runs over the SM fermions $f\in\{\ell,q\}$, the neutrino flavours $\alpha,\beta\in\{e,\mu,\tau\}$, and the chiralities $P\in\{P_L, P_R\}$. In general, the off-diagonal coefficients $\varepsilon^{fP}_{\alpha\beta}$ can be complex and introduce additional CP-violation in the neutrino sector. 

Since we are interested in the solar neutrino signal, we are mostly concerned with neutrino interactions with ordinary matter, i.e.~with the first generation charged fermions  $f\in\{e, u, d\}$. If we further assume that the neutrino flavour structure of the NSI is independent of the charged fermion sector, we can factorise the NSI coefficient as~\cite{Esteban:2018ppq}
\begin{equation}\label{eq:nsi_def}
 \varepsilon^{fP}_{\alpha\beta} \equiv \varepsilon^{\eta,\varphi}_{\alpha\beta}\,\xi^{fP}\,,
\end{equation}
where $\varepsilon^{\eta,\varphi}_{\alpha\beta}$ is the overall magnitude of the NSI 
and $\xi^{fP}$ describes the relative coupling strength with the fermion $f$. In this notation, we can respectively define vector and axial-vector interactions as 
\begin{equation}
	\varepsilon^f_{\alpha\beta} \equiv \varepsilon^{fL}_{\alpha\beta} + \varepsilon^{fR}_{\alpha\beta} \equiv \varepsilon^{\eta,\varphi}_{\alpha\beta}\,\xi^{f} \quad \text{and} \quad
	\tilde\varepsilon^f_{\alpha\beta} \equiv \varepsilon^{fL}_{\alpha\beta} - \varepsilon^{fR}_{\alpha\beta} \equiv \tilde\varepsilon^{\eta,\varphi}_{\alpha\beta}\,\tilde\xi^{f} \,, 
\end{equation}
with $\xi^f \equiv \xi^{fL} + \xi^{fR}$ and $\tilde{\xi}^f \equiv \xi^{fL} - \xi^{fR}$. Since neutrino matter propagation is mostly affected by vector interactions, we will be most interested in the vector NSI $\varepsilon^f_{\alpha\beta}$ in this work and assume any axial-vector NSI to be negligible $\tilde\varepsilon^f_{\alpha\beta}\ll\varepsilon^f_{\alpha\beta}$.

To describe both propagation and scattering effects induced by NSI, it is convenient to consider the effective interactions with the building blocks of stable matter, namely the electron, proton, and neutron. 
Following the conventions introduced in our previous work~\citep{Amaral:2023tbs}, we will work in the basis $\{\varepsilon^e_{\alpha\beta}, \varepsilon^p_{\alpha\beta},\varepsilon^n_{\alpha\beta}\}$, where 
\begin{equation}
\begin{split}
    \varepsilon^p_{\alpha\beta} = 2 \varepsilon^u_{\alpha\beta} + \varepsilon^d_{\alpha\beta} \quad \text{and} \quad
    \varepsilon^n_{\alpha\beta} = \varepsilon^u_{\alpha\beta} +  2\varepsilon^d_{\alpha\beta} \,.
\end{split}
\label{eq:eps_p_n}
\end{equation}

The corresponding relative NSI strengths are given by\footnote{Note the peculiar normalisation factor, $\sqrt{5}$, which was originally introduced in Ref.\cite{Esteban:2018ppq} to have unit vectors $\xi^u$ and $\xi^d$, and which we keep for consistency with the literature.} 
\begin{equation}
    \xi^e \equiv\sqrt{5} \cos \eta \sin \varphi, \quad \xi^p \equiv \sqrt{5} \cos \eta \cos \varphi, \quad \xi^n \equiv \sqrt{5} \sin \eta\,.
    \label{eq:xi_projections}
\end{equation}
The two angles $\eta$ and $\varphi$ represent the relative strength of the proton-to-neutron and proton-to-electron NSI, respectively. The choice of $\varphi = 0$ corresponds to the nucleon NSI plane, while $\eta = 0$ corresponds to the charged NSI plane. We refer to Sec.~II.1 of Ref.~\cite{Amaral:2023tbs} for a detailed description and a visualisation of the parametrisation.

\subsection{Neutrino Propagation}

The first ingredient we need to compute the neutrino scattering rate in~\cref{eq:trace} is the density matrix $\rho$ of the neutrinos arriving at the detector after propagating from the Sun to the detector.
NSI affect the matter potential experienced by neutrinos as they travel through a dense medium. Their evolution is governed by the total Hamiltonian in matter, which, when assuming a neutral medium, can be expressed in the flavour basis as
\begin{equation}
H_m = U_\mrm{PMNS} \frac{1}{2 E_\nu}\left(\begin{matrix}
0 & 0 & 0 \\
0 & \Delta m_{21}^{2} & 0 \\
0 & 0 & \Delta m_{31}^{2}
\end{matrix}\right) U_\mrm{PMNS}^{\dagger}+V(x)
\left(\begin{matrix}
1 + \mathcal{E}_{ee}(x) & \mathcal{E}_{e\mu}(x) & \mathcal{E}_{e\tau}(x) \\
\mathcal{E}_{e\mu}^*(x) & \mathcal{E}_{\mu\mu}(x) & \mathcal{E}_{\mu\tau}(x) \\
\mathcal{E}_{e\tau}^*(x) & \mathcal{E}_{\mu\tau}^*(x) & \mathcal{E}_{\tau\tau}(x) \\
\end{matrix}
\right)\,,
\label{eq:hamiltonian_nsi}
\end{equation}
where $U_\mrm{PMNS}$ is the PMNS matrix as written in Eq.~(9) of Ref.~\cite{Amaral:2023tbs}, $\Delta m_{ij}^2 \equiv m_i^2 - m_j^2$, $V(x)=\sqrt{2} G_F N_e(x)$, and $N_{e}(x)$ is the spatially-dependent electron number density. The effective NSI parameters $\mathcal{E}_{\alpha\beta}$ are defined as
\begin{equation}\label{epss}
    \mathcal{E}_{\alpha\beta}(x) \equiv \sum_{f} \frac{N_f(x)}{N_e(x)} \varepsilon_{\alpha \beta}^f=\left[\left(\xi^p + \xi^e\right)+Y_{n}(x) \xi^{n}\right] \varepsilon_{\alpha \beta}^{\eta, \varphi}= \xi(x) \varepsilon_{\alpha \beta}^{\eta, \varphi}\,,
\end{equation}
where we have used the fact that in neutral matter $N_p(x) = N_e(x)$ and defined the neutron-to-electron number ratio as $Y_n(x) \equiv N_n(x)/N_e(x)$~\cite{Bahcall:2005va}.
To compute the $S$-matrix for propagation, and eventually the density matrix $\rho$ from it,  we must solve the time-dependent Schr\"odinger equation governed by the matter Hamiltonian in \cref{eq:hamiltonian_nsi} for the propagating neutrino states:
\begin{equation}\label{eq:SE_gen}
    i \, \partial_x     \begin{pmatrix}
     \nu_e \\
     \nu_\mu \\
     \nu_\tau
    \end{pmatrix} = H_m  \begin{pmatrix}
     \nu_e \\
     \nu_\mu \\
     \nu_\tau
    \end{pmatrix} \,.
\end{equation}
On their way to underground direct detection experiments, the solar neutrinos propagate first through the Sun and then through the Earth.

\subsubsection{Solar Neutrino Propagation}
\label{sec:solar_nu}

For the propagation of neutrinos through the solar medium, we can reduce the full three-flavour problem described by the $3 \times 3$ Hamiltonian of~\cref{eq:hamiltonian_nsi} into an effective two-flavour one. This is warranted as the third mass eigenstate effectively decouples in the Sun since both $\Delta m_{31}^2 \gg \Delta m_{21}^2$~\cite{Esteban:2024eli} and $\Delta m_{31}^2 \gg 2 E_\nu V_{\mathrm{cc}}=2 E_\nu \sqrt{2} G_F N_e(x)$, rendering the Hamiltonian effectively block-diagonal. 
Furthermore, NSI with $|\varepsilon_{\alpha\beta}|\lesssim3$ satisfy the additional condition that $G_F \sum_f N_f(x)\varepsilon_{\alpha\beta}^f \ll \Delta m_{31}^2 / (2E_\nu)$, which ensures that the extra matter effects induced by NSI do not spoil this decoupling~\cite{Amaral:2023tbs}.

Rotating from the conventional (electroweak charged-current) neutrino flavour basis to a new basis $\bm{\hat\nu} =R_{13}^\dagger R_{23}^\dagger\, \bm{\nu} $, with $R_{13}$ and $R_{23}$ the conventional rotation matrices about the neutrino mixing angles $\theta_{13}$ and $\theta_{23}$, we can write the 
neutrino evolution equation in the full $3\times 3$ picture
as
\begin{equation}\label{eq:nu_evol}
    i\, \partial_x
    \begin{pmatrix}
    \hat \nu_e \\
    \hat \nu_\mu \\
    \hat \nu_\tau
    \end{pmatrix}
    = 
    \begin{pmatrix}
    H^\mathrm{eff} && 0 \\
    0 && \frac{\Delta m^2_{31}}{2\, E_\nu} 
    \end{pmatrix} 
    \begin{pmatrix}
    \hat \nu_e \\
    \hat \nu_\mu \\
    \hat \nu_\tau
    \end{pmatrix} \,,
\end{equation}
where $H^\mathrm{eff}$ denotes the effective $2\times2$ matter mixing Hamiltonian.
In order to evolve the neutrino states  through the solar medium, we diagonalise this effective $2\times2$ Hamiltonian and solve the time-dependent Schr\"odinger equation step-by-step along the neutrino path through the solar interior. 
Following the conventions of Ref.~\cite{Esteban:2018ppq}, this effective Hamiltonian reads
\begin{equation}\label{eq:ham_eff}
H^\mathrm{eff} =  \frac{\Delta m_{21}^2}{4 E_{\nu}} \left(\begin{matrix}
-\cos 2 \theta_{12} & e^{i\dCP}\sin 2 \theta_{12}   \\
e^{-i\dCP}\sin 2 \theta_{12} & \cos 2 \theta_{12}
\end{matrix}\right) + V(x)\left[\left(\begin{matrix}
c_{13}^{2} & 0 \\
0 & 0
\end{matrix}\right)+\xi(x)\left(\begin{matrix}
-\varepsilon_{D}^{\eta, \varphi} & \varepsilon_{N}^{\eta, \varphi} \\
(\varepsilon_{N}^{\eta, \varphi})^* & \varepsilon_{D}^{\eta, \varphi}
\end{matrix}\right)\right]\,,
\end{equation}
with the coefficients $\varepsilon_{N}^{\eta, \varphi}$ and $\varepsilon_{D}^{\eta, \varphi}$ being linear combinations of $\varepsilon_{\alpha \beta}^{\eta, \varphi}$. We give explicit expressions for $\varepsilon_{N}^{\eta, \varphi}$ and $\varepsilon_{D}^{\eta, \varphi}$   in~\cref{eq:nsi_eff}.

In the following, we outline the diagonalisation of the Hamiltonian and the derivation of the full evolution $S$-matrix,
while we refer the reader to~\cref{app:solarnu} for further details.
We diagonalise the Hamiltonian by a unitary transformation  $U^{m\dagger}_{12} H^{\mathrm{eff}} U^m_{12} = \operatorname{diag}(E^m_{1}, E^m_{2})$, where the transformation matrix is given by 
\begin{equation}\label{eq:Urot}
 U^m_{12} (x)=\begin{pmatrix} \cos \theta_{12}^m & e^{i\chi}\sin \theta_{12}^m \\ -e^{-i\chi}\sin \theta_{12}^m & \cos \theta_{12}^m \end{pmatrix} \,,
\end{equation}
with the matter mixing angle $\theta_{12}^m$ and an effective CP-violating phase $\chi$ given by
\begin{equation}\label{tanchi}
 \tan \chi =  -  \frac{  (\Delta m_{21}^2/4 E_{\nu})  \sin 2\theta_{12} \sin \dCP +V(x)\, \xi(x)\, \text{Im}(\varepsilon_N)}{ (\Delta m_{21}^2/4 E_{\nu})  \sin 2\theta_{12} \cos \dCP + V(x)\, \xi(x)\, \text{Re}(\varepsilon_N)} \,.
\end{equation}
Note that this phase depends on both $\dCP$ and the imaginary part of the NSI coefficient $\varepsilon_N$. 
Hence, this effective phase $\chi$ is only present if either $\dCP\neq 0$ and/or the NSI elements are complex.
For the expressions of the matter energy eigenvalues $E_1^m$ and $E_2^m$, as well as the matter mixing angle $\theta_{12}^m$, we refer to~\cref{eq:pq,eq:en_diff,eq:mat_energies,eq:mat_angles}.

Assuming adiabatic evolution of the neutrinos through the Sun, we can solve~\cref{eq:nu_evol} approximately analytically by integrating from the neutrino production point $x_0$ to the solar surface. For this approach to be valid, we must ensure that the adiabatic  approximation holds along the entire neutrino trajectory through the Sun, requiring the diagonalised Hamiltonian $H^\mathrm{eff}$ to remain approximately diagonal when evolving the neutrino states from a point $x$ to an infinitesimally close point $x+\Delta x$ . We can then compute the adiabaticity parameter $\gamma$, defined in~\cref{eq:def_adiab}, which compares the relative magnitude of the diagonal elements of $H^\mathrm{eff}$ to the off-diagonal elements. 

We study this assumption in detail in~\cref{app:solarnu} and investigate the impact of NSI and a non-zero $\dCP$ on adiabaticity in~\cref{app:adiab}. 
We find that the adiabatic approximation holds even when including extra NSI-induced matter effects.
This significantly simplifies the full evolution of neutrinos through the solar medium and allows us to compute the full solar neutrino evolution $S$-matrix in the mass basis as
\begin{equation}\label{eq:Smatrixfull}
    S =
    \begin{pmatrix}
   e^{i\phi} && 0 && 0 \\
   0 && e^{-i\phi} && 0 \\
    0 && 0 && e^{-i\Phi_{33}}
    \end{pmatrix} \,
    U^m_\mathrm{PMNS}(x_0)^\dagger\,.
\end{equation}
The phase $\phi \equiv \int_{0}^L \Delta E_{21}^m(x)\, \dd x $ represents the integrated phase of the upper diagonalised $2\times2$ block, while $\Phi_{33} \equiv  (\Delta m^2_{31} / 2 E_\nu) L$ is the integrated phase of the decoupled third mass eigenstate. The matrix $U^m_{\rm PMNS}(x_0)=R_{23}\,R_{13}\,U^m_{12}(x_0)$ is the PMNS matrix in matter with the matter mixing angle $\theta_{12}^m$ evaluated at the point of neutrino creation $x_0$. Once the neutrinos exit the Sun, they free-stream to Earth as vacuum mass eigenstates, where they decohere on their journey.
A comprehensive derivation of the intermediate evolution matrices, including the non-adiabatic factors, is detailed in \cref{app:solarnu}.

From the $S$-matrix in~\cref{eq:Smatrixfull}, we can derive the three-flavour density matrix for solar neutrinos reaching the Earth, $\rho^{(e)} = S \pi^{(e)} S^\dagger$, where $\pi^{(e)}=\mathrm{diag}(1,0,0)$ is the projector onto the initial electron-neutrino state.
The expression for the neutrino density matrix at the Earth is further simplified by two key physical effects during propagation. The first is due to \textit{spatial averaging}. Since solar neutrinos are produced across a finite volume within the Sun, the initial matter mixing angle $\theta_{12}^m(x)$ is position dependent. We account for this by computing the following expectation value for each neutrino population $p$, with $p \in \{pp, \mrm{^8 B}, \dots\}$:
\begin{equation}
    \langle \cos^2(\theta_{12}^{m}) \rangle_p = \int_{0}^{1}\cos^2(\theta_{12}^{m}(x))\, f_p(x)\,dx\,.
\end{equation}
Here, $x \equiv r/R_\odot$ is the fractional solar radius and $f_p(x)$ represents the spatial production distribution for each neutrino population $p$, which we take from the Standard Solar Model predictions given in Ref.~\cite{Vinyoles:2016djt}. The second effect is the \textit{decoherence}. The macroscopic propagation distance from the Sun to the Earth, coupled with the inherent uncertainty in the exact neutrino production point, forces the oscillatory phase factors (such as $e^{i\phi}$ and $e^{i\Phi_{33}}$) to average to zero over many oscillation periods. Because of this complete decoherence, all off-diagonal elements in the mass-basis density matrix vanish. Hence, the solar neutrinos arrive at Earth as an incoherent superposition of mass eigenstates, described by the diagonal density matrix
\begin{equation}\label{eq:density_mass}
    \rho_{m}^\odot = 
    \begin{pmatrix}
    c_{13}^2\ \langle c_m^2\rangle && 0 && 0 \\
   0 && c_{13}^2\ \langle s_m^2\rangle && 0 \\
    0 && 0 && s_{13}^2
    \end{pmatrix}\,.
\end{equation}

Finally, since we are ultimately interested in the scattering of solar neutrinos, and because for neutral-current NSI the flavour eigenstates coincide with the electroweak-interaction eigenstates, we can retrieve the interaction-base neutrino density matrix from the mass-base density matrix via a transformation with the vacuum PMNS matrix,
\begin{equation}\label{eq:rho_solar}
    \rho^\odot = U_\mathrm{PMNS}\ \rho_{m}^\odot \ U_\mathrm{PMNS}^\dagger\,.
\end{equation}
For reference, we give the explicit expressions for the elements of $\rho^\odot$ in the interaction basis in~\cref{eq:rho_flav}.

\subsubsection{Earth Propagation}
\label{sec:earth_matter}

To account for the full journey of solar neutrinos to the detector,
we must incorporate matter effects from the propagation of the neutrinos through the interior of the Earth into the density matrix $\rho^\odot$ in~\cref{eq:rho_solar}. 
In \snudd, we do this by implementing the slab method, where a neutrino's path through a medium is approximated as going through a series of small, constant density slabs~\cite{Lisi:1997yc,Akhmedov:1998ui,Chizhov:1999he,Ohlsson:1999um,Akhmedov:2006hb, Giunti:2007ry,Akhmedov:2008qt,Blennow:2013rca, Maltoni:2015kca,Kelly:2021jfs, Gonzalo:2023mdh, Denton:2025oxf}. Where traditionally analytic expressions have been obtained using a small number of slabs~\cite{Lisi:1997yc,Akhmedov:1998ui,Chizhov:1999he,Ohlsson:1999um}, we perform the numerical evaluation with a high number ($\usim 50$) of slabs. This method has the advantage of being fully applicable to different Earth models. 

Within a slab $k$, we evaluate the average matter Hamiltonian $\langle H_m\rangle_k$ from \cref{eq:hamiltonian_nsi}. We then diagonalise this $3\times3$ matrix numerically to give $H_m^{\rm diag}=\mathrm{diag}(E^1_m, E^2_m, E^3_m)$,  such that the propagation equation given in \cref{eq:SE_gen} can be solved straightforwardly. The transformation matrix $U^m$ that diagonalises $H_m$ is also required to calculate the $S$-matrix in the flavour basis: 
\begin{equation}
    S_k = U_k^{m}
      \begin{pmatrix}
      e^{-iE_1^m\,\Delta x_k} & 0 & 0 \\
      0 & e^{-iE_2^m\,\Delta x_k} & 0 \\
      0 & 0 & e^{-iE_3^m\,\Delta x_k}
      \end{pmatrix}
      U_k^{m\,\dagger},
\end{equation}
where we have added the subscript $k$ to signify the relevant quantities for the $k^{\rm th}$ slab, with a given width $\Delta x_k$. To determine the full Earth propagation effect, we consider the product of all slabs, such that
\begin{equation}
    S(E_\nu,\eta_\mrm{nad}) = \prod_{k=1}^{N_{\rm st}} S_k,
\end{equation}
where we have explicitly written the dependence on neutrino energy, which is explicit in~\cref{eq:hamiltonian_nsi}, and $\eta_\mrm{nad}$ is the nadir angle, which fixes the neutrino path once the detector location and time is fixed, as detailed in \cref{app:Earth_motion}.

To evaluate the average Hamiltonian in a given slab $\langle H_m\rangle_k$, we must determine the electron number density $N_{e}(x)$ across it. Typically, the Earth matter density is modelled using PREM~\cite{Dziewonski:1981xy}, which describes the Earth as a series of spherical shells, each parametrised by a cubic polynomial in radius. The electron fraction $Y_e$ is taken to be a piecewise constant, $Y_e \simeq 0.466$ in the core and $Y_e \simeq 0.494$ in the mantle, with $R_{\rm mantle}=0.546R_{\oplus}$ and $R_\oplus \approx 6371\,\mrm{km}$ the radius of the Earth. The effective electron number density is $N_e = Y_e\, \rho_m /m_{\rm mol}$, with $m_{\rm mol}$ the molar mass of Earth material and $\rho_m$ its density.

In~\cref{app:nutrajEarth}, we detail the geometry of the neutrino trajectory through the Earth and how we relate that to the radial coordinates in  which PREM is defined. We have followed Ref.~\cite{Lisi:1997yc} and defined $\eta_\mrm{nad}$ as the angle between the upwards direction from the detector and the incident neutrino. 
There are a number of other codes that also compute neutrino propagation effects, such as \texttt{NuSQuIDS}~\cite{Arguelles:2021twb}, \texttt{NuFAST}~\cite{Denton:2024pzc,Denton:2025oxf}, and \texttt{PEANUTS}~\cite{Gonzalo:2023mdh}.
We have compared our Earth matter evolution to \texttt{NuFAST}~\cite{Denton:2024pzc,Denton:2025oxf} assuming SM interactions and achieved good agreement.

To accelerate the above computation in \snudd, we have also implemented a fast analytical approximation based on the Magnus expansion~\cite{Magnus:1954zz,DOlivo:1990cwf,Ioannisian:2008ve,Supanitsky:2008eq}.
The Magnus expansion allows us to compute an approximate solution to a quantum mechanical differential equation in terms of a series expansion. 
In \snudd, we use this to compute the leading order expansion of the evolution $S$-matrix as the solution of the Schr\"odinger equation in~\cref{eq:SE_gen}. At high energies where matter effects are non-perturbatively large, the full numerical slab method is preferred. 

Finally, we note that Earth-matter effects become mostly relevant only for the highest energy neutrinos of the solar $^8$B and $hep$ fluxes (see~\cref{fig:probs_diag}). Therefore, given the additional computational cost of calculating the Earth-matter effects, we advise users to only use these routines when calculating neutrino fluxes with energies about $\sim 5\,{\rm MeV}$.

\subsection{NSI Cross Sections}\label{ernr}

As well as modifying neutrino propagation, neutrino NSI also affect the interaction cross-section within the detector. To provide a comprehensive description of solar neutrino detection, we consider two distinct channels that direct detection experiments are sensitive to: \cevns and \eves.

\paragraph{\cevns cross section:}

To compute the generalised \cevns cross,  we require the square of the amplitude summed over all possible final neutrino flavours $\gamma$:
\begin{equation}
\begin{split}
    \left(\diff{\zeta_{\nu N}}{E_R}\right)_{\alpha\beta} &  = \frac{G_F^2\, m_N}{\pi}\left(1-\frac{M_N\, E_R}{2 E_\nu^2}\right)\\
    &\times \sum_\gamma\ \langle gs|| \hat G^\mathrm{SM}\,\delta_{\alpha\gamma} + (\hat G^\mathrm{NSI}_{\alpha\gamma})^* ||gs\rangle \langle gs|| \hat G^{\mathrm{SM}}\,\delta_{\gamma\beta} + \hat G^{\mathrm{NSI}}_{\gamma\beta} ||gs\rangle\\
    & = \frac{G_F^2\, M_N}{\pi}\left(1-\frac{m_N\, E_R}{2 E_\nu^2}\right)\ F^2(E_R) 
    \\ &  \times
    \left[ \frac{1}{4} \, Q_{\nu N}^2\, \delta_{\alpha\beta} - Q_{\nu N} \, \operatorname{Re}(G^\mathrm{NSI}_{\alpha\beta}) + \sum_\gamma (G^\mathrm{NSI}_{\alpha\gamma})^*G^{\mathrm{NSI}}_{\gamma\beta}\right]\, ,
    \label{eq:sig_CEVNS_gen}
\end{split}
\end{equation}
where $F(E_R)$ is the Helm form factor~\cite{Helm:1956zz}, and  ${Q_{\nu N} = N - (1 -4\,\sin^2\theta_W)\,Z}$ is the SM coherence factor. Furthermore, we have used the Hermiticity of the NSI nucleus coupling, defined by
\begin{align}\label{eq:g_cevns}
    G^\mathrm{NSI}_{\alpha\beta} &\equiv \left(2\,\varepsilon^{u}_{\alpha\beta} + \varepsilon^{d}_{\alpha\beta} \right) Z +  \left(\varepsilon^{u}_{\alpha\beta} + 2\,\varepsilon^{d}_{\alpha\beta} \right) N= (\xi^p\, Z + \xi^n \, N)\ \varepsilon_{\alpha\beta}^{\eta,\varphi} \,.
\end{align}

\paragraph{\eves cross section:}

First, we compute the cross section for $\nu - e$ scattering. In the presence of NSI, there are three separate contributions: the SM  $Z$- and $W$-exchange, as well as the NSI Fermi-like contact term. Summing the amplitudes and integrating over phase space, we find the expression for the generalised neutrino-electron scattering cross section in the presence of complex NSI,

\vspace{-0.3cm}

\begin{equation}\label{eq:xsec_el}
\begin{split}
        \left(\diff{\zeta_{\nu e}}{E_R}\right)_{\alpha\beta} =  \frac{2 \, G_F^2\, m_e}{\pi}\ \sum_\gamma\,\Bigg\{&(G^L_{\alpha\gamma})^*G^{L}_{\gamma\beta} + (G^R_{\alpha\gamma})^* G^{R}_{\gamma\beta}\left(1-\frac{E_R}{E_\nu}\right)^2 \\
        - \big[&(G^L_{\alpha\gamma})^* G^{R}_{\gamma\beta} + (G^R_{\alpha\gamma})^* G^{L}_{\gamma\beta} \big] \frac{m_e\, E_R}{2 E_\nu^2} \Bigg\} \,,    
\end{split}
\end{equation}
where we have defined the generalised neutrino-electron couplings as
\begin{equation}
    G^P_{\alpha\beta} \equiv
    g^e_{P\alpha}\delta_{\alpha\beta} + \varepsilon^{eP}_{\alpha\beta} \quad \text{and}   \quad  (G^P_{\alpha\beta})^* \equiv
    g^e_{P\alpha}\delta_{\alpha\beta} + (\varepsilon^{eP}_{\alpha\beta})^*\,,
\end{equation}
with $P\in\{L,R\}$ denoting the chirality of the scattered electron. 
For vector NSI, we have $\varepsilon_L=\varepsilon_R$.
The SM electroweak neutrino-electron couplings are given in terms of the $Z$-coupling $g^f_P = T^3_f - \sin^2\theta_w\, Q^\mathrm{EM}_f$ as
\begin{equation}
    g^e_{P\alpha} = \begin{cases}  1 + g^e_L  & \text{if}\ \alpha=e \ \text{and} \ P=L \,, \\
    g^e_P  & \text{otherwise} \,.
    \end{cases} 
\end{equation}

The cross section in \cref{eq:xsec_el} captures the free-electron cross section in the absence of any potential well. However, in reality, neutrinos at direct detection experiments interact with either atomically bound electrons in noble element detectors or valence band electrons that must be promoted to the higher energy conduction band in semiconductor targets. These binding effects can have a significant impact on the expected event rate, potentially reducing it by $\usim 20\%\text{--}30\%$~\cite{Chen:2016eab}. 

Since we focus on xenon-based detectors, we consider the former effect, which we model via a stepping approximation that weighs the free-electron cross section by the number of orbital electrons that can be freed with a particular energy deposit. Concretely, we take
\begin{equation}
    \left(\diff{\zeta_{\nu e}}{E_R}\right)_{\alpha\beta} \rightarrow \frac{1}{Z} \sum_{i=1}^Z \Theta(E_R - B_i) \left(\diff{\zeta_{\nu e}}{E_R}\right)_{\alpha \beta}\,,
    \label{eq:stepping}
\end{equation}
where $\Theta$ is the Heaviside step function and $B_i$ is the binding energy of the $i^\mrm{th}$ electron~\cite{Kopeikin:1997abc}. A more precise modelling of atomic ionization is more complex. Currently, most experimental collaborations rely on the results from Ref.~\cite{Chen:2016eab}, which models xenon's atomic response using the relativistic random phase approximation (RRPA) based on \textit{ab initio} calculations~\cite{PhysRevA.20.964,PhysRevA.20.978}. The resulting spectra is reduced by approximately 20\% from stepping approximation.

\section{Presenting \texttt{SNuDD}}
\label{sec:snudd}

We present \snudd (Solar Neutrinos for 
Direct Detection): a Python-based, open-source codebase for accurately computing solar neutrino scattering rates due to neutrino-electron and -nucleus scattering. \snudd can be used to compute the SM scattering rates and the modified rates in the presence of BSM physics, such as in the form of NSI. \snudd has been developed to consistently incorporate BSM neutrino physics effects both in neutrino propagation (solar and terrestrial) and in neutrino scattering within the detector. Additionally, \snudd provides functionality to incorporate detector effects like energy thresholds, selection efficiencies, and resolution effects for generating realistic signal spectra.

Below, we describe the main functionality of \snudd: from specifying an NSI model, to computing the neutrino density matrix, to calculating the scattering rate with experimental effects included. The code is freely available
under an open-source license at
\href{https://github.com/SNuDD/SNuDD}
{\faGithub\ \texttt{github.com/SNuDD/SNuDD}}. All code examples shown in this section are bundled in a Jupyter notebook named 
\texttt{quick\_start.ipynb}
that comes with \snudd and can be run interactively.

\subsection{Setting Up a Model}

First, we set up an NSI model. To specify a model, we must provide the $3\times3$ matrix of the NSI magnitudes in flavour space $\varepsilon^{\eta,\varphi}_{\alpha\beta}$, as well as the relative NSI strength with electrons, protons, and neutrons, which is encoded in the angles $\eta$ and $\varphi$, as described in~\cref{sec:theory}.

In the following example code, we show how to set up an NSI model instance for a purely off-diagonal $\mu\tau$-coupling of magnitude $\nsi{\mu\tau}=0.1$, equal coupling strengths with protons and neutrons ($\eta=\pi/4$), and no coupling with electrons ($\varphi=0$).

\begin{lstlisting}
from snudd.models import GeneralNSI

# Define NSI parameters
NSI_matrix = numpy.array([[0.0, 0.0, 0.0],
                          [0.0, 0.0, 0.1],
                          [0.0, 0.1, 0.0]])
NSI_eta = numpy.pi/4
NSI_phi = 0

# Create NSI model object
NSI_model = GeneralNSI(NSI_matrix, NSI_eta, NSI_phi)
\end{lstlisting}

\subsection{Calculating the Density Matrix}

Next, we demonstrate how to compute the solar neutrino density matrix, both for \textit{day-time} (only solar matter evolution) and \textit{night-time} observation (combined solar + Earth-matter evolution). We note that the base unit of energy in \snudd is GeV. Hence, care should be taken when specifying neutrino energies and recoil energies.

The following lines of code demonstrate how to setup a \texttt{DensityMatrixCalculator} object and calculate the density matrix elements over a range of neutrino energies. This simple call of the \texttt{density} method will only compute the \textit{day-time} density matrix, in this case for the $^8\mrm{B}$ neutrinos.
\begin{lstlisting}
from snudd.nsi.nsi_probabilities import DensityMatrixCalculator

# Create probability calculator 
NSI_density_calculator = DensityMatrixCalculator(NSI_model)

# Calculate solar density matrix over neutrino energy range
Enus = numpy.geomspace(1e-6, 1e0, 200) # in GeV
solar_density = NSI_density_calculator.density(Enus, '8B')
\end{lstlisting}

\paragraph{Including Earth-matter Effects}
We can include earth-matter effects in the solar neutrino propagation. For this, we need to specify a detector location and track the solar neutrino incident angles (or equivalently their paths through Earth's interior) over the experimental data taking period. \snudd provides the \texttt{SolarAngles} module, allowing the user to compute a weighted histogram of neutrino incident angles (outputting $\cos\eta_{\rm nad}$) over a given data taking period at a specified detector location.

\begin{figure}[t!]
\centering
\includegraphics{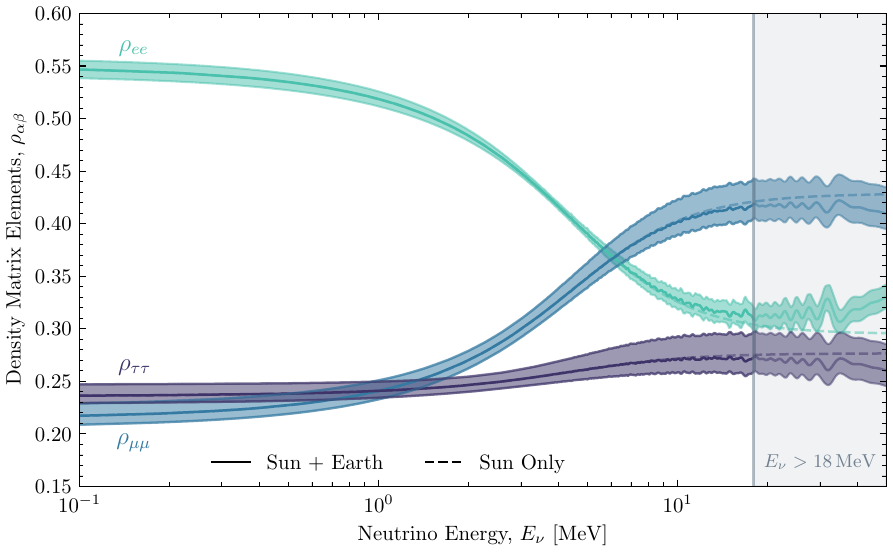}
\caption{The diagonal elements of the density matrix with neutrino energy $E_\nu$, corresponding to the electron survival and transition probabilities, computed using \snudd. Solid lines show the result of propagating neutrinos through both the Sun and the Earth, while dashed lines show the result for solar propagation only. While  energies above $E_\nu \gtrsim 18\,\mrm{MeV}$ are inaccessible by solar neutrinos, we extend our results to higher energies to better illustrate the Earth-matter effects, which become more important at higher energies. The bands capture the $1\sigma$ uncertainties in the density matrix elements due to the uncertainties in oscillation parameters.}
\label{fig:probs_diag}
\end{figure}

In the following example, we show how to generate the histogram of incident angles for the Gran Sasso location (XENONnT) for a data taking campaign of one year:

\begin{lstlisting}
from snudd.geometry import SolarAngles

# Define experimental parameters (location, data taking period)
lat_gs = 42.47  # Latitude of Gran Sasso in degrees north
t0     = 0      # t0 - beginning in days after 3rd January (perihelion)
T      = 365    # T  - duration of data taking in days

# Define solar angles calculator for Gran Sasso
GranSasso = SolarAngles(latitude=lat_gs, t0=t0, T=T)

# Generate histogram of cos(nadir) 
cnadirs, weights = GranSasso.cnadir_hist(bins=30)
\end{lstlisting}
With this weighted histogram of neutrino incident angles, we can compute the full neutrino density matrix by averaging the Earth-matter effects over the incident angles using the \texttt{DensityMatrixCalculator} as follows:
\begin{lstlisting}
# Calculate full density matrix over neutrino energy range
Enus = numpy.geomspace(1e-6, 1e0, 200) # in GeV
earth_density = NSI_density_calculator.density_earth(Enus, cnadirs, weights, nu='8B')
\end{lstlisting}

We show the diagonal elements of the density matrix $\rho_{\alpha\alpha}$ generated by \snudd in \cref{fig:probs_diag}. The element $\rho_{ee}$ is equivalent to the electron neutrino survival probability, while $\rho_{\mu\mu}$ and $\rho_{\tau\tau}$ correspond to the transition probabilities into muon and tau neutrinos, respectively. 
Below the maximum solar neutrino energy of $18\,\mrm{MeV}$, associated with $hep$ neutrinos, the matrix elements are dominated by propagation effects in the Sun. While energies above this are not relevant for our work, we extend our results into this regime to illustrate that Earth-matter effects become important at higher energies. We also compute the uncertainties in these elements by first re-computing the density matrix elements assuming the extremal $1\sigma$ values for each of the oscillation parameters given in Ref.~\cite{Esteban:2024eli}. We then calculate the resulting uncertainties due to the variation in each of the oscillation parameters by taking the difference with the best-fit results. Finally, we sum these contributions in in quadrature to give us the total uncertainty in the elements.

\subsection{Computing the Neutrino Recoil Spectrum}
\label{sec:spectrum}

In \snudd, the full pipeline for computing 
the neutrino recoil spectrum via the trace formalism in~\cref{eq:trace}
is implemented in the \texttt{Target} class. More precisely, its two subclasses
\texttt{Nucleus}
and
\texttt{Electron}
enable the user to define scattering target objects for calculating nuclear and electron recoil event rates, respectively. 

\medskip

\noindent\textbf{\cevns spectrum:}
The following code block illustrates how to generate a \cevns spectrum.
To do so, we first have to specify the nuclear scattering target via a \texttt{Nucleus} object. We then have to update it with our current \texttt{GeneralNSI} model, which holds the expressions for the relevant \cevns cross section. Next, we pre-compute the neutrino density matrix via the call \texttt{prepare\_density} including the Earth-matter evolution. Note that simply calling \texttt{prepare\_density()} without arguments computes only the \textit{day-time} density without any Earth-matter effects.
Finally, we compute the \cevns spectrum over a predefined range of recoil energies (specified in GeV).

\begin{lstlisting}
from snudd import config
from snudd.targets import Nucleus

# Create scattering target
Xe_nucleus = Nucleus(54, 132, mass=131.9041535 * config.u) # single isotope 

# Load NSI model in scattering target and generate density matrix
Xe_nucleus.update_model(NSI_model)
Xe_nucleus.prepare_density(cnadirs=cnadirs, cnadir_weights=weights)

# Compute recoil spectrum
E_Rs = numpy.geomspace(1e-2, 1e2, 500) / 1e6  # Recoil energy in GeV
NSI_spec_nr = Xe_nucleus.spectrum(E_Rs)
\end{lstlisting}

\medskip

\noindent\textbf{\eves spectrum:}
Computing accurate \eves spectra requires knowledge of the available electrons for scattering at a given neutrino energy. \snudd's \texttt{Electron} object requires information about the host nucleus and its respective orbital binding energies. For a generic atom, the user can calculate the resulting electron recoil spectra according to the stepping approximation \cref{eq:stepping}. For xenon targets, one can instead produce a spectrum scaled according to the RRPA~\cite{Chen:2016eab}. The handling of the binding energies in the appropriate scattering prescription is implemented in the \texttt{Electron} class, and we illustrate how to generate an electron recoil spectrum in the following block of code.

\begin{figure}
    \centering
    \includegraphics{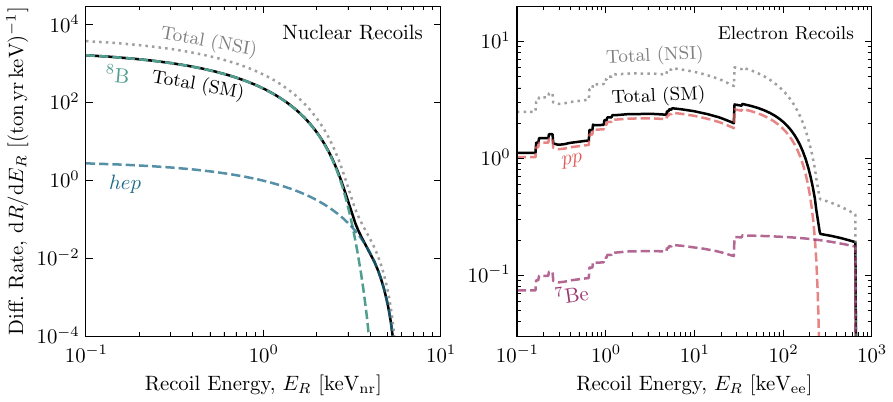}
    \caption{The differential neutrino scattering rate~\cref{eq:trace} for a $\mathrm{^{132}Xe}$ detector before detector-dependent effects are folded in for the Standard Model (SM) and a choice of non-standard interactions (NSI). \textbf{Left:} The nuclear recoil spectrum. The two dominant contributions arise from the $\mathrm{^8B}$ and $hep$ fluxes. The chosen NSI model has $\nsi{ee}=1$ and $\eta=\varphi=0$. \textbf{Right:} The electron recoil spectrum. The two dominant contributions arise from the $pp$ and $\mathrm{^7Be_{864\,keV}}$ fluxes. The free-electron rate is modified by a step function, modelling the binding energies of electron orbitals within a xenon nucleus and becoming important below energies $35\,\mrm{keV_{ee}}$, and an energy-dependent scaling between  $0.25\,\mrm{keV_{ee}}\text{--}30\,\mrm{keV_{ee}}$  to account for the relativistic random phase approximation (RRPA)~\cite{Johnson:1979aaa,Huang:1979aaa,Huang:1981oux}. The chosen NSI model has $\nsi{ee}=0.3$, $\eta=0$, and $\varphi=\pi/2$.}
    \label{fig:rates}
\end{figure}

As for \cevns, we first setup a \texttt{Nucleus} object for the host atom, which we use together with a \texttt{binding} object holding the orbital binding energies and an \texttt{rrpa\_scaling} object to initialize an \texttt{Electron} object. As before, we a to feed the \texttt{Electron} object to current \texttt{GeneralNSI} model and pre-compute the neutrino density matrix. Finally, we can generate the \eves spectrum over a range of recoil energies.

\begin{lstlisting}
from snudd import config
from snudd.targets import Nucleus, Electron   # still requires the host nucleus
from snudd.binding import binding_xe          # dataclass of binding energy data of xenon 
from snudd.rrpa    import rrpa_scaling        # rrpa scaling for bound electron

# Create host nucleus
Xe_nucleus = Nucleus(54, 132, mass=131.9041535 * config.u) # single isotope 

# Create bound electron object
Xe_electron = Electron(Xe_nucleus, binding_xe, rrpa_scaling) 
Xe_electron.update_model(NSI_model)
Xe_electron.prepare_density()

# Compute recoil spectrum
E_Rs = numpy.geomspace(1e-2, 1e2, 500) / 1e6  # Recoil energy in GeV
NSI_spec_er = Xe_electron.spectrum(E_Rs)
\end{lstlisting}

In~\cref{fig:rates}, we show example \cevns and \eves spectra for neutrino scattering in xenon produced with \snudd. The left panel shows the total differential \cevns rate, both for the SM (black solid) and an NSI of $\nsi{ee}=1$ with protons (grey dotted). We further show the breakdown of the total SM rate into the relative contributions from the $^8\mrm{B}$ (green dashed) and $hep$ (blue dashed) solar neutrino fluxes, which dominate the \cevns spectrum.
In the right panel, we show the corresponding \eves recoil spectrum both for the SM (black solid) and an NSI of $\nsi{ee}=0.3$ with electrons (grey dotted). We also show the breakdown of the SM rate into the relative contributions from the $pp$ (red dashed) and 
$^7\mrm{Be}$ (purple dashed) solar neutrino fluxes, which dominate the \eves spectrum.

\subsection{Including Detector Effects}

To make realistic predictions of neutrino scattering rates at direct detection experiments, we apply
an energy-dependent efficiency $\epsilon(E)$ to the theoretical recoil spectrum from~\cref{sec:spectrum} and convolve it with a detector resolution response function $\Phi$.
\snudd provides the framework for this bundled into the \texttt{Convolver} class, and contains pre-defined efficiency and resolution functions for LZ, XENONnT, and PandaX-4T. 

In the following example code, we illustrate how to compute the experimental count rate for solar neutrino-electron scattering in a future xenon experiment modelled after XLZD or PandaX-xT.  We first set up a \texttt{Convolver} object and initialize it with the recoil \texttt{spectrum} 
evaluated over a range of energies (\texttt{E\_Rs}), an \texttt{efficiency} function, and a \texttt{resolution} function. To obtain realistic predictions, we employ the latest LZ efficiency function~\cite{xu_2025_17660135} and energy resolution reported in Ref.~\cite{LUX:2016rfb}. By calling the method \texttt{convolved\_binned\_rate(E1, E2)}, we compute the convolved number of signal counts in an energy bin with bin edges $[E_1, E_2]$. Applying this prescription over a predefined set of energy bins, we can generate the experimental binned histogram of \eves events.

\begin{lstlisting}
from snudd.efficiencies import efficiency_lz_er_WS24
from snudd.resolution   import res_lz_er
from snudd.resolution   import Convolver

# Counts @ future xenon detector
def counts_future_xe(specturm, E_Rs, E1, E2):
    exposure_future = 200. # t yrs
    convolution_lz_sig = Convolver(E_Rs, specturm, efficiency_lz_er_WS24, res_lz_er)
    return convolution_lz_sig.convolved_binned_rate(E1, E2) * exposure_future

# Definine histogram binning 
bin_edges   = numpy.linspace(1, 20, 20, endpoint=True) / 1e6 # Bin edges in GeV 
bin_width   = bin_edges[1] - bin_edges[0]
bin_centers = bin_edges[:-1] + bin_width/2

# Apply detector effects to neutrino spectrum
counts_er = [counts_future_xe(NSI_spec_er, E_Rs, (bc - bin_width/2), (bc + bin_width/2)) for bc in bin_centers]
\end{lstlisting}

\begin{figure}
    \centering
    \includegraphics{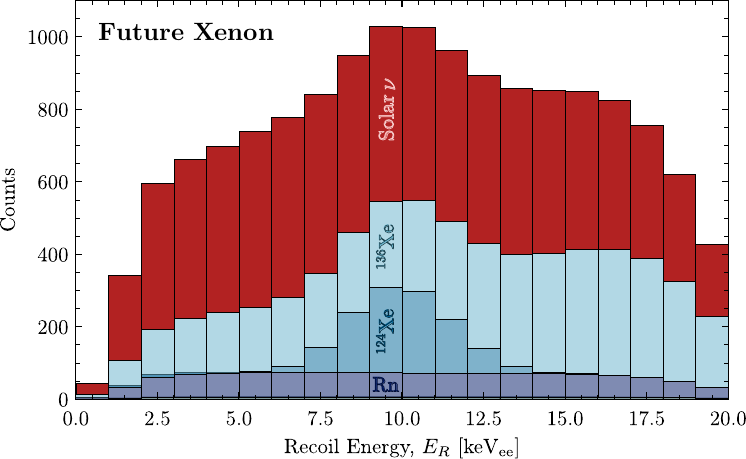}
    \caption{Stacked histogram of the electron recoil counts expected at a future xenon detector due to solar neutrinos, as computed by \snudd, and background events. The backgrounds include krypton (shown as the bottom-most bars but not labelled), radon, the spectral peak arising from the double electron capture from $^{124}\mrm{Xe}$ (which is smeared by resolution effects), and the double $\beta$-decay associated with $^{136}\mrm{Xe}$. The detector is based on the $200$ \tnyr exposure of the planned XLZD and PandaX-xT experiments, with backgrounds taken from Ref.~\cite{DARWIN:2020bnc}.}
    \label{fig:ER_hist}
\end{figure}

The example code above generates the binned histogram of electron recoil events induced by solar neutrinos (red)  shown  in~\cref{fig:ER_hist}, together with the forecasted backgrounds for a future xenon detector.

\section{Application to Xenon-based Direct Detection Experiments}
\label{sec:application}

To demonstrate the practical application of \snudd, we use it to derive new constraints on the neutrino NSI parameters with direct detection. Motivated by the recent observation of solar neutrinos in xenon-based experiments, we focus on the TPC detectors LZ, XENONnT, and PandaX-4T. We also project the reach of a future xenon experiment based on XLZD and PandaX-xT.  

To constrain NSI parameters, we must compute the number of events in a set number of recoil energy bins. For this, the differential scattering rates must be mapped to the observable energy window of xenon detectors. The physically observed differential recoil rate is obtained by convolving the predicted spectra in \cref{fig:rates} with a detector's energy-dependent efficiency profile $\epsilon$ and resolution response function $\Phi$. The observed differential rate is then 
\begin{equation}
    \label{eq:dr_dd_obs}
    \diff{R}{E_R}(E_R) = \int_0^\infty \diff{R}{E_R'}(E_R')\,\epsilon(E_R')\, \Phi\textbf{(}E_R, E_R', \sigma(E_R')\textbf{)} \dd E_R'\,,
\end{equation}
where $\dd R / \dd E_R'$ is the theoretical  differential rate computed via \cref{eq:trace}. Throughout this analysis, we take $\Phi$ to be a Gaussian centred at the observable recoil energy $E_R$ with a detector-dependent standard deviation $\sigma$ characterising its energy resolution. 
We compute the number of expected neutrino events within the $i^\mrm{th}$ energy bin $N_\nu^i$ by integrating the convolved rate within the energy range of the bin $[E^i_1,\, E^i_2]$ and performing an isotopic sum over the relative abundances $X_A$ of the xenon targets. The count is thus given by
\begin{equation}
    \label{eq:n_dd}
    N^i_\nu = \mathcal{E} \sum_A X_A \int_{E^i_1}^{E^i_2} \diff{R_A}{E_R}\, \dd  E_R\,,
\end{equation}
where $\mathcal{E}$ is the total experimental exposure and $\dd R_A / \dd E_R$ is the differential rate for a given isotope $A$ computed via~\cref{eq:dr_dd_obs}. This count and our statistical analysis are treated differently depending on whether we are focusing on nuclear or electron recoils. 

\medskip

\noindent\textbf{Nuclear Recoils:}
The nuclear recoil channel is sensitive to solar \cevns, which primarily arises from the high-energy flux of $\mrm{^8B}$ neutrinos occurring near the threshold energy of xenon detectors. Given the localized nature of this signal and the fact that backgrounds can be efficiently suppressed or neglected in this narrow low-energy region, we consider only a single bin, yielding the total number of counts. Since backgrounds are suppressed, we assume the dominant uncertainty in our NR analysis to arise from the $12\%$ uncertainty in the $\mrm{^8B}$ neutrino flux~\cite{Vinyoles:2016djt}. Moreover, as detector resolutions $\sigma$ are usually reported in terms of electron-equivalent energies, we first convert our computed \cevns differential rate and efficiency functions into electron-equivalent energies using a quenching factor. We use the Lindhard quenching factor~\cite{Lindhard1963INTEGRALEG}; further details can be found in Ref.~\cite{Amaral:2023tbs}.

\medskip

\noindent\textbf{Electron Recoils:}
The electron recoil channel probes solar \eves, which is dominated by the low-energy solar $pp$ neutrino flux. Within the typical energy window that ERs are analysed ($\usim2\,\mrm{keV_{ee}}\text{--}30\,\mrm{keV_{ee}}$), the backgrounds are multi-component and can be highly spectral. A single-bin approach is therefore inadequate; instead, we employ a multi-bin analysis. Since the ER background is not negligible, we take the dominant systematic uncertainty to arise from the reported uncertainties in them for the different data sets we consider. For our future projection, we instead assume that the backgrounds will be under greater control and rather take this uncertainty to be dominated by that in the $pp$ neutrino flux of $1\%$~\cite{Vinyoles:2016djt}.

\subsection{Current and Future Liquid Xenon Observatories}
\label{sec:exps_details}

Here, we list and detail the experimental configurations we use to perform our analysis, all of which are xenon dual-phase time projection chambers (TPCs).

\begin{itemize}

\item{\bf LZ:}
The LZ experiment \cite{Mount:2017qzi, AKERIB2020163047} operates in the Sanford Underground Research Facility (SURF) in Lead, South Dakota, USA, with a 4300 meter water-equivalent overburden to shield from cosmic radiation. The detector consists of 7 tonnes of liquid xenon. For the nuclear recoil analysis, we utilize the data and efficiency curve reported in Ref.~\cite{LZ:2025igz} with an exposure of $5.7$ \tnyr, adopting the energy resolution from Ref.~\cite{LUX:2016rfb}. The efficiency curve reaches $50\%$ of its maximum at $1.88\,{\rm keV}_{\rm nr}$. We assume a background-free hypothesis for this analysis. We have validated this setup by reproducing a consistent number of $^8$B neutrino events (20.2 events).

For the electronic recoil analysis, we employ the 2022 and 2024 datasets \cite{LZ:2022lsv, LZ:2024zvo} which correspond to exposures of $0.9$ \tnyr and $3.3$ \tnyr respectively. Our analysis is based on the 1D ROI analysis reported in Ref.~\cite{LZ:2025zpw}, we take the efficiency for both data sets from the data provided~\cite{xu_2025_17660135}. To validate the detector modelling, we compared our SM expected counts to those reported by the LZ collaboration; we find good agreement for the 2022 dataset ($35.2$ vs $38.1\pm2.3$) and reasonable agreement for the 2024 dataset ($132.4$ vs $151\pm9$). Furthermore, we find that the inclusion of RRPA is necessary to avoid overestimating the predicted counts, which increased to $163.0$ when the approximation was disabled. For backgrounds we take those reported in Ref.~\cite{LZ:2025zpw} (excluding solar neutrinos) and assume the background uncertainty scales as the quadrature sum of the component backgrounds. This approach is justified as the 2024 analysis statistics are dominated by flat backgrounds. 

\item{\bf XENONnT:}
Located at the INFN Laboratori Nazionali del Gran Sasso (LNGS) in Italy, the XENONnT experiment \cite{XENON:2024wpa} is a dual-phase TPC containing 5.9 tonnes of liquid xenon as active target. For the nuclear recoil analysis, we adopt the data and efficiency curves reported in Refs.~\cite{XENON:2026ydt,dacheng_xu_2026_20576156}, with an exposure of $6.77$ \tnyr.  The efficiency curve reaches $50\%$ of its maximum at $2.21\,{\rm keV}_{\rm nr}$. We assume a background-free hypothesis and utilize the energy resolution function from Ref.~\cite{XENON:2020rca}. This analysis is validated by the observation of $19.7$ $^8$B neutrino events, consistent with the expectation of $16^{+5}_{-4}$.

For the electronic recoil analysis, we follow the methodology described in Ref.~\cite{Amaral:2023tbs}, utilizing the efficiency, data, and background fits from Ref.~\cite{XENON:2022ltv} for an exposure of $1.16$ \tnyr. To account for the energy-dependent uncertainties of the background components, we restrict our analysis to the $\left[0-30\right]\,{\rm keV}$ region; in this interval, the background is dominated by $^{214}$Pb with an associated uncertainty of $12.5\%$. The validation of the detector modelling for ERs are discussed in detail in Ref.~\cite{Amaral:2023tbs}.

\item{\bf PandaX-4T:}
The PandaX-4T dark matter direct search experiment is located in the China Jinping Underground Laboratory (CJPL) \cite{PandaX:2022aac} and utilizes a dual-phase xenon TPC with a sensitive volume of 3.7 tonnes of liquid xenon. For the nuclear recoil analysis, we utilize the efficiency function and the paired scintillation and ionization dataset reported in Ref.~\cite{PandaX:2024muv} with an exposure of $1.20$ \tnyr and adopt the energy resolution function from Ref.~\cite{PandaX:2022ood}.  The efficiency curve reaches $50\%$ of its maximum at $1.61\,{\rm keV}_{\rm nr}$. We assume a background-free hypothesis, which is validated by the observation of 3.2 $^8$B neutrino events, consistent with the 2-hit expectation presented in Table II of Ref.~\cite{PandaX:2024muv}.

For the electronic recoil analysis, we use the data, efficiency, and backgrounds from Ref.~\cite{PandaX:2024cic}. We restrict our analysis to the Run 1 dataset, corresponding to an exposure of $363.3$ tonne-days, as the Run 0 dataset is dominated by the tritium background. The associated relative background uncertainty, calculated in quadrature, is  $6.2\%$. We validate the detector modelling against the collaboration expectation of $72.6\pm8.1$ events; we find 69 events without RRPA and 56 with it.

\item{\bf Future Xenon:}
Currently, there are two future xenon-based detectors in the preparatory phase: XLZD \cite{XLZD:2024nsu} and PandaX-xT~\cite{PANDA-X:2024dlo}. XLZD is a merger of the XENON, LUX, and DARWIN collaborations, whereas PandaX-xT is the successor to the PandaX-4T detector. Both represent the ultimate evolution of liquid-xenon TPC technology. Since both proposed detectors feature a baseline exposure of 200 \tnyr and  the specific detector capabilities are not yet settled, we forecast the discovery and exclusion potential of a generic xenon 200 \tnyr detector as a proxy for both. For the NR projection, we assume no backgrounds and model the detector as essentially a 200 \tnyr version of XENONnT. We conservatively use the same efficiency curve as XENONnT, though this may improve in the development of a future experiment; it should be understood that modest improvements in the threshold energy for NR analysis are anticipated. Note that in our previous work, we augmented the XENON efficiency curve reported in Ref.~\cite{XENON:2018voc}
to reach $\usim 50\%$ at $3\,{\rm keV_{\rm nr}}$, which reaches a maximum of $\sim0.8$ (compared to $\sim0.16$ of the current XENONnT~\cite{XENON:2026ydt}). As we highlighted in that work, any substantial improvement in the threshold would have to be accompanied by an improvement in the $^8$B uncertainty, as it is this that dominates the statistical uncertainty and limits our sensitivity. As our data, we use an Asimov data set, assuming that we observe data exactly consistent with the expectation value of the neutrino count under the SM hypothesis~\cite{Cowan:2010js}.

For ERs, we take the efficiency function from the more recent LZ study in Ref.~\cite{LZ:2025zpw} and, for consistency, we take the LZ resolution function from Ref.~\cite{LUX:2016rfb}. We assume the projected reduced backgrounds presented in Ref.~\cite{DARWIN:2020bnc}. As in the NR case, we use an Asimov data set, assuming SM counts and background expectations within each energy bin. In comparison with our previous work in Ref.~\cite{Amaral:2023tbs}, we have an additional background from $^{124}$Xe, which reflects the difference in expected backgrounds from Ref.~\cite{Baudis:2013qla} to Ref.\cite{DARWIN:2020bnc}.

\end{itemize}

\subsection{Statistical Framework}

We use the standard profile log-likelihood ratio test described in  Ref.~\cite{Amaral:2023tbs} for inference. The likelihood function 
is constructed from the product of a Poisson term, which accounts for the binned event counts, and Gaussian constraint terms designed to capture experimental uncertainties via nuisance pull parameters:
\begin{equation}
\begin{split}
    \mathcal{L}(\nsi{\alpha\beta},\eta, \varphi; a, b) &\equiv \prod_i^{N_\mathrm{bins}} \mathrm{Po}\left[N^i_\mathrm{obs}\big|(1 + a) N^i_\nu(\nsi{\alpha\beta}, \eta, \varphi) + (1 + b) N^i_\mathrm{bkg} \right] \\ 
    &\times \mathrm{Gauss}\left(a \big| 0, \sigma_{a}\right)
    \mathrm{Gauss}\left(b \big| 0, \sigma_{b}\right)\,.
\end{split}
\label{likelihood}
\end{equation}
Here, $N^i_\mathrm{obs}$ is the number of observed neutrino events, $N^i_\nu$ is the number of expected neutrino events as computed via \cref{eq:dr_dd_obs}, and $N^i_\mathrm{bkg}$ is the number background events, all of which fall in the $i^\mathrm{th}$ energy bin. The pull parameters $a$ and $b$ govern the effect of uncertainties in the neutrino flux and detector backgrounds, with standard deviations $\sigma_a$ and $\sigma_b$, respectively. In practice, we assume that one of these pull parameters dominates at a time. For our NR analysis, we take the effect of $a$ to dominate, setting $\sigma_a = 0.12$, reflecting the uncertainty in the $\mrm{^8 B}$ neutrino flux. Conversely, for our ER analysis, we assume the effect of $b$ dominates when setting our limits, taking $\sigma_b$ to be the fractional background uncertainty values we quote in our above experimental descriptions. For the future projections, we instead again assume that $a$ dominates and $\sigma_a = 0.01$, in line with the $pp$ neutrino flux uncertainty.

To set limits on $\nsi{\alpha\beta}$, we employ a profile log-likelihood ratio test using the two-sided test statistic
\begin{equation}
    \label{eq:test_stat}
    t_\varepsilon \equiv -2 \ln\left[\frac{\mathcal{L}(\nsi{\alpha\beta}, \eta=\eta_0, \varphi=\varphi_0;\hat{a},\, \hat{b})}{\mathcal{L}(\hat{\hat{\varepsilon}}^{\eta, \varphi}_{\alpha\beta}, \eta=\eta_0, \varphi=\varphi_0; \hat{\hat{a}},\,\hat{\hat{b}})}\right]\,.
\end{equation}
In the above, single hats denote the maximum likelihood
estimators of the likelihood conditioned on the value of $\nsi{\alpha\beta}$ being tested, while double hats represent the maximum likelihood estimators of the unconstrained likelihood. 
We perform a raster scan over the angles $\eta$ and $\varphi$, computing $t_\varepsilon$ in~\cref{eq:test_stat} with $\eta_0$ and $\varphi_0$  fixed at each grid point.
Assuming Wilks' theorem holds asymptotically in the high-statistics limit, $t_\varepsilon$ follows a $\chi^2$-distribution with one degree of freedom. Consequently, we set the $90\%$ Confidence Level (CL) upper bounds at the critical threshold $t_\varepsilon^\mathrm{lim} = 2.71$~\cite{ParticleDataGroup:2024cfk}.

\subsection{Results}

In our analysis, we consider only one NSI parameter at a time and derive our sensitivities in the nucleon ($\varphi = 0$) and charged ($\eta = 0$) NSI planes. We take the best-fit values for the oscillation parameters from the NuFIT 6.1 results, which has $\dCP=212^\circ$~\cite{Esteban_2024}. This is in contrast to our previous work, where we set $\delta_{\rm CP}=0^\circ$. As discussed in~\cref{app:adiab}, a non-zero $\dCP$ alters the adiabaticity parameter~\cref{eq:def_adiab}, and the particular choice of the NuFIT best fit value of $\dCP=212^\circ$ makes the adiabatic approximation valid for the parameter space of interest.  
Moreover, adopting this value instead of $\dCP = 0^\circ$ effectively interchanges the
transition probabilities from electron neutrinos to muon  and tau neutrinos, $\rho_{\mu\mu}$ and  $\rho_{\tau\tau}$.
This can be seen explicitly by comparing the probabilities shown in~\cref{fig:probs_diag} (for which we set $\dCP=212^\circ$) to Fig.~2.9 of Ref.~\cite{Amaral:2023nmz} (where $\dCP = 0^\circ$).
It also shifts the direct detection sensitivities to muon- and tau-flavoured NSI elements. 
Since muon and tau neutrinos cannot be differentiated with either \cevns or \eves,
this leads to an interchange of the sensitivities of $\nsi{\mu\mu}\leftrightarrow\nsi{\tau\tau}$ and $\nsi{e\mu}\leftrightarrow\nsi{e\tau}$ compared to our previous results in Ref.~\cite{Amaral:2023tbs}.

\subsubsection{Nucleon NSI Plane}
\label{sec:res_nuc}

\begin{figure}[]
    \begin{center}
    \includegraphics[]{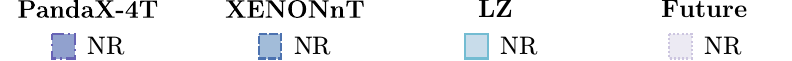}

    \vspace*{-0ex}
    \includegraphics[width=0.47\textwidth]{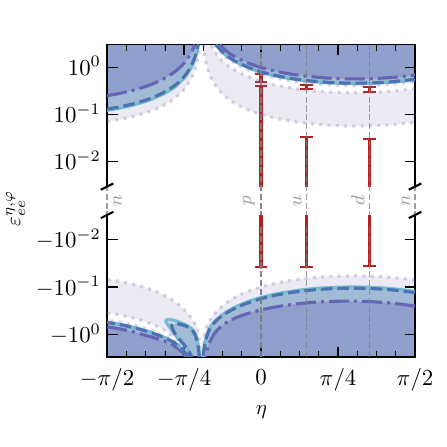}
    \includegraphics[width=0.47\textwidth]{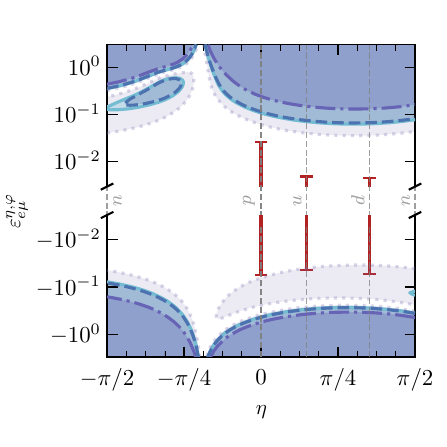}

    \vspace*{-6ex}
    \includegraphics[width=0.47\textwidth]{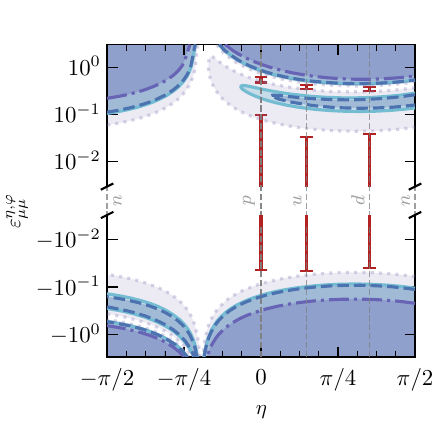}
    \includegraphics[width=0.47\textwidth]{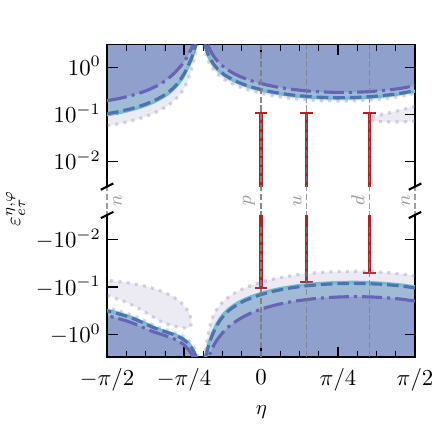}

    \vspace*{-6ex}
    \includegraphics[width=0.47\textwidth]{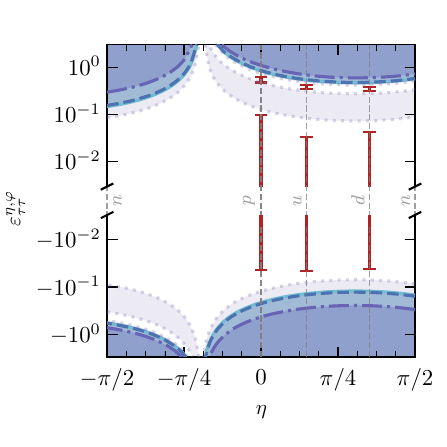}
    \includegraphics[width=0.47\textwidth]{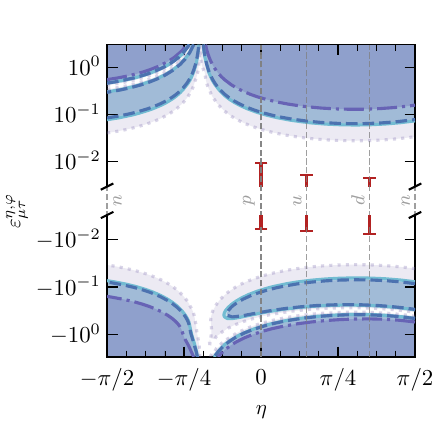}
    \end{center}

    \vspace{-3.5ex}
    \caption{The 90\% CL exclusion limits on the NSI parameters $\varepsilon^{\eta,\varphi}_{\alpha\beta}$ as a function of the angle $\eta$ in the nucleon NSI plane. The six panels show the constraints for the distinct flavour combinations. We plot the limits derived from the PandaX-4T, XENONnT, and LZ experiments, as well as the projected performance of a future xenon detector based on XLZD and PandaX-xT. The shaded areas indicate the parameter space excluded by each experiment. We also show the allowed ranges from global fits for NSI with protons~\cite{Coloma:2019mbs} and quarks~\cite{Coloma:2023ixt}.}
  \label{fig:limits_eta}
\end{figure}

We present the $90\%$ CL~exclusion limits and projected sensitivities of our analysis in the nucleon NSI plane ($\varphi=0$) in~\cref{fig:limits_eta}. From less constraining (smaller excluded areas) to more constraining (larger excluded areas), we show the derived limits from PandaX-4T (dot-dashed contours), XENONnT (dashed contours), and LZ (solid contours), as well as the future sensitivity of a future xenon detector based on XLZD and PandaX-xT (dotted contours).
This hierarchy in the strength of the limits is due to the combination of the respective exposure and detection efficiency at low energies. While XENONnT has a larger exposure ($6.77$ \tnyr) than LZ ($5.7$ \tnyr), LZ has a better detection efficiency at low energies than XENONnT and therefore yields slightly better limits. PandaX-4T has a much smaller exposure ($1.20$ \tnyr) and therefore is least constraining.

Our results are to be compared with the NSI limits established by the global analyses of Ref.~\cite{Coloma:2019mbs} (proton NSI) and Ref.~\cite{Coloma:2023ixt} (up- and down-quark NSI), 
which combined COHERENT data alongside various oscillation experiments.  The globally allowed values for the NSI parameters are represented by the red bars and are shown for three standard directions in the NSI parameter space: pure couplings to protons ($\eta=0$), up-quarks ($\eta=\tan^{-1}(1/2)$), and down-quarks ($\eta=\tan^{-1}(2)$). 
Crucially, we see that direct detection constraints are now comparable in reach. Indeed, for the diagonal NSI elements $\varepsilon_{\alpha\alpha}$ and the off-diagonal element $\varepsilon_{e\tau}$, the leading constraints from LZ and XENONnT begin to overlap with the allowed intervals from global fits.  This strongly motivates the inclusion of direct detection data in future global analyses of NSI.

To help compare with global-fit results~\cite{Coloma:2019mbs,Coloma:2023ixt} and the results of specific experimental efforts, we tabulate the 90\% CL intervals of our PandaX-4T, XENONnT, and LZ \cevns analyses for   $\varepsilon^u_{\alpha\beta}$, $\varepsilon^d_{\alpha\beta}$, $\varepsilon^p_{\alpha\beta}$, and
$\varepsilon^n_{\alpha\beta}$ in~\cref{tab:nr_results}.
Note that our analysis yields slightly more constraining intervals than those reported recently in the \cevns analysis of Ref.~\cite{DeRomeri:2026prc}; however, this is expected due to several differences in our analyses. First, we consider a 1D strategy in which we allow for only one non-zero NSI element $\nsi{\alpha\beta}$ at a time, while Ref.~\cite{DeRomeri:2026prc} marginalises over all NSI elements. Second,
in our formalism, we take into account the full quantum-mechanical expectation value of the \cevns scattering cross section via the trace over the neutrino density matrix, shown in~\cref{eq:trace}. This is necessary to capture the neutrino flavour correlations and the resulting cancellation effects correctly for off-diagonal NSI elements, as pointed out in Ref.~\cite{Amaral:2023tbs}.

Our limits shown in~\cref{fig:limits_eta} 
exhibit various interesting features. For instance, \cevns analyses of NSI suffer from a material-dependent blind direction. This originates from the accidental cancellation of the coherence factor in the NSI \cevns coupling in~\cref{eq:g_cevns}~\cite{Scholberg:2005qs,Barranco:2005yy,Dent:2017mpr,Dutta:2020che}. Specifically, in a particular NSI direction $\eta$, the proton- and neutron-NSI contributions cancel out, resulting in $\xi^p\, Z + \xi^n \, N \rightarrow0$. This is the same effect that results in blind directions in the search for dark matter~\cite{Feng:2013vod,Cheek:2023zhv}, and for NSI this occurs when 
\begin{equation}
    \eta=\tan ^{-1}\left(-\frac{Z}{N} \cos \varphi\right)\,,
\end{equation}
where $Z$ and $N$ are the number of protons and neutrons for the target isotopes, respectively. Since liquid xenon is not mono-isotopic but rather a mixture of naturally occurring isotopes, this cancellation is not exact. The ratio of $Z/N$ is very similar for most heavy elements including the abundant xenon isotopes in particular. Hence, the location of the blind direction in a xenon experiment is approximately at $\eta\approx - \pi/5$ in the nucleon NSI plane ($\varphi=0$). Detectors employing targets of different heavy elements will suffer from a blind direction at a very similar location in the nucleon NSI plane.
This, for example, can be seen in the recent NSI sensitivity study using archaeological lead in the RES-NOVA experiment Ref.~\cite{RES-NOVA:2026fii}. As was pointed out in Ref.~\cite{Amaral:2023tbs}, combining data with experiments using light elements as scattering targets instead could significantly alleviate this blind direction due to their very different $Z/N$ ratio.

Apart from this fixed, material-dependent blind direction, there are also insensitivity bands across the entire $\eta$ domain, leading to gaps in the limits. These bands are also due to interference effects in which NSI effects cancel. In the simplest of cases, when only one diagonal NSI element $\varepsilon_{\alpha\alpha}$ is non-zero, the cancellation can be understood at the level of the cross section given in~\cref{eq:sig_CEVNS_gen}. In this case, only the cross section element $\left(\dd{\zeta_{\nu N}}/\dd{E_R}\right)_{\alpha\alpha}$ will be non-zero, and hence the cancellation simply occurs between the SM-NSI interference term and quadratic NSI term, resulting in the cancellation condition~\cite{Dutta:2020che}
\begin{equation}
    \varepsilon^{\eta,\varphi}_{\alpha\alpha} = \frac{Q_{\nu N}}{\xi^p\, Z + \xi^n \, N}\,.
\end{equation}
The location of this insensitivity band is also dependent on the target material through $Z$ and $N$. 
This further strengthens the motivation to combine data from xenon-based detectors with 
experiments employing light targets with sufficiently different $Z$ and $N$.

For off-diagonal NSI $\varepsilon_{\alpha\beta}$, with $\alpha\neq \beta$, the cancellation is more intricate. In this case, due to the flavour sum in the quadratic NSI term of~\cref{eq:sig_CEVNS_gen}, the $\alpha\beta$-, $\alpha\alpha$- and $\beta\beta$-elements of the cross section all contribute and must cancel with the appropriate factors of the density matrix in the trace. For a detailed discussion of the interference structure in~\cevns at direct detection experiments, see Sec.~IV.2 of Ref.~\cite{Amaral:2023tbs}, where the analytical conditions for cancellation are derived and their phenomenological implications are discussed.

\begin{figure}[]
    \begin{center}
    \includegraphics[]{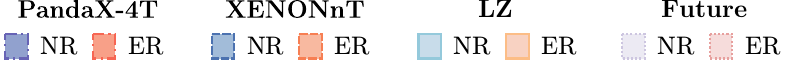}

    \vspace*{-0ex}
    \includegraphics[width=0.47\textwidth]{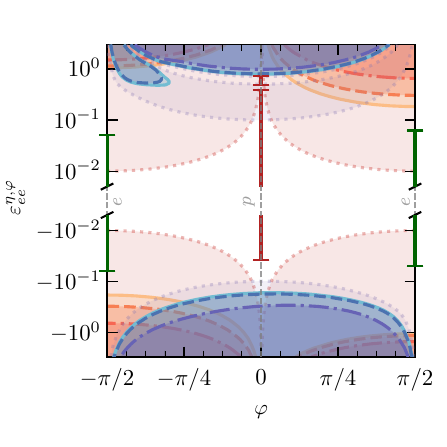}
    \includegraphics[width=0.47\textwidth]{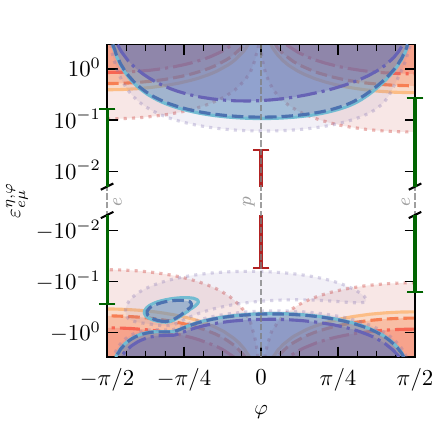}

    \vspace*{-6ex}
    \includegraphics[width=0.47\textwidth]{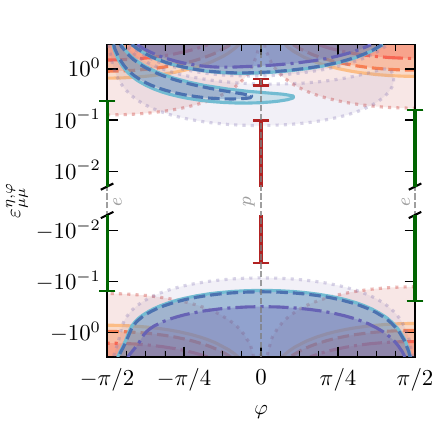}
    \includegraphics[width=0.47\textwidth]{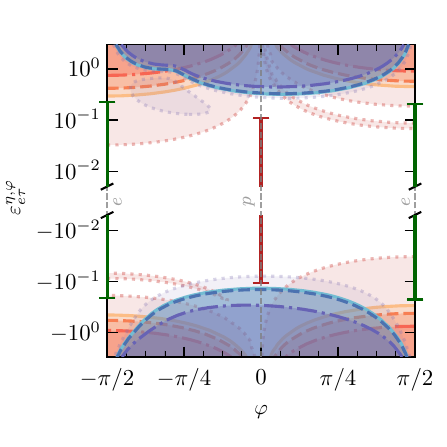}

    \vspace*{-6ex}
    \includegraphics[width=0.47\textwidth]{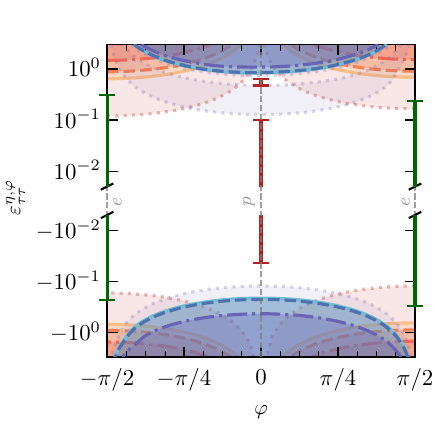}
    \includegraphics[width=0.47\textwidth]{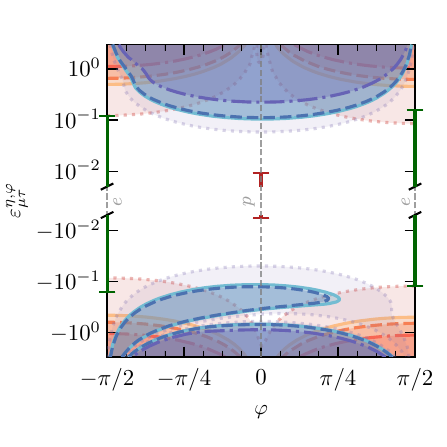}
    \end{center}

    \vspace{-3.5ex}
    \caption{Same as in \cref{fig:limits_eta} but for varying $\varphi$ in the charged NSI plane ($\eta = 0$). Both NR and ER constraints are overlayed for all six distinct flavour combinations. Also shown are the allowed ranges from global fit results for NSI with protons (red bars)~\cite{Coloma:2019mbs} and electrons (green bars)~\cite{Coloma:2023ixt}.}
  \label{fig:limits_phi}
\end{figure}

\subsubsection{Charged NSI Plane}

\begin{figure}
    \includegraphics[]{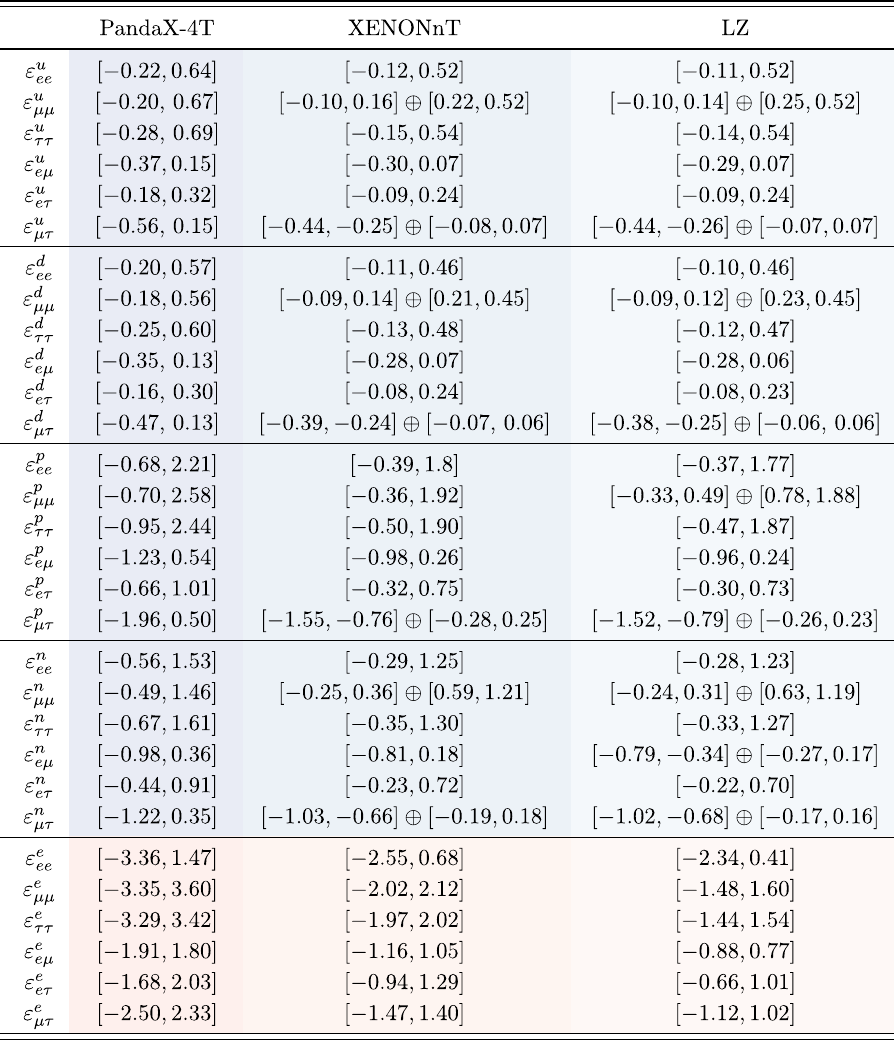}
    \captionsetup{type=table}
    \caption{The 
    $90\%$ CL allowed intervals for NSI in the up-quark, down-quark, proton, neutron and electron directions. The limits quoted here have been converted from the NSI magnitude $\nsi{\alpha\beta}$ to the more familiar Lagrangian parameters $\varepsilon^f_{\alpha\beta}$ via~\cref{eq:nsi_def} and the relations~\cref{eq:eps_p_n,eq:xi_projections}.  Shown are the results from our NR and ER analysis of the Pandax-4T, XENONnT, and LZ experiments.}
    \label{tab:nr_results}
\end{figure}

To completely explore the charged NSI plane ($\eta=0$), we must analyse both \cevns and \eves at direct detection experiments since the charged contribution can arise from both the protons in nuclei and the atomic electrons.
In~\cref{fig:limits_phi}, we collect the  combined limits from both the NR and the ER analyses of direct detection data in the charged NSI plane. As  in~\cref{fig:limits_eta}, blue colours represent NR limits, while orange colours correspond to ER limits. The contour styles match those from our NR analysis above. 
For ER, the hierarchy of the limit strength is rooted in the respective exposures and  backgrounds of the different experiments. LZ has the largest ER data set (4.2 \tnyr), followed by XENONnT (1.16 \tnyr) and PandaX-4T ($363.3$ tonne-days).
In~\cref{fig:limits_phi}, the complementarity of NR and ER for NSI studies is manifest. While NR searches are most constraining for $\varphi=0$ (pure proton NSI), they lose sensitivity towards the boundaries 
for $\varphi\to\pm\pi/2$ (pure electron NSI). Conversely, searches employing ER attain the maximum sensitivity at $\varphi=\pm\pi/2$, while they lose their sensitivity for $\varphi\to0$ (pure proton NSI). Combining  both channels provides optimal sensitivity in the charged NSI plane, fully covering the blind directions in $\varphi$ discussed above.
The capability to simultaneously observe \cevns and \eves via the NR and ER channel makes direct detection experiments ideal for probing NSI.

In~\cref{fig:limits_phi}, we also show the allowed intervals from global fits in the proton direction ($\varphi=0$, red bar)~\cite{Coloma:2019mbs} and electron direction ($\varphi=\pm\pi/2$, green bars)~\cite{Coloma:2023ixt} for comparison. 
We highlight the complementary sensitivity of ERs in direct detection and note that these
are now reaching sensitivities comparable to the global fit intervals for $\varepsilon_{e\mu}$ and $\varepsilon_{e\tau}$. 
Moreover, our sensitivity projections show that ER searches at 
future planned direct detection experiments, such as XLZD and PandaX-xT, will be able to improve significantly upon current limits, potentially improving the constraints on $\nsi{ee}$, $\nsi{e\mu}$, and $\nsi{e\tau}$ by approximately an order of magnitude beyond the currently allowed global-fit values. 
In~\cref{tab:nr_results}, we tabulate the 90\% CL intervals from our ER analysis of direct detection data for pure electron-NSI, $\varepsilon^e_{\alpha\beta}$ (orange rows).
In particular, our analysis based on the recent LZ data set~\cite{LZ:2025zpw} improves over previous NSI searches with ER data at direct detection experiments~\cite{Amaral:2023tbs,Celestino-Ramirez:2025snn}.

In~\cref{fig:limits_phi}, we also observe insensitivity bands for the $\nsi{ee}$ and $\nsi{e\tau}$ elements.
Analogously to the NR case, these insensitivity bands arise due to accidental cancellations in the NSI-dependent terms of the rate. 
Beyond the visible gap in the projected sensitivity to $\nsi{e\tau}$, there is a fine-tuned region of the parameter space in $\nsi{ee}$ where the SM rate is recovered for non-zero NSI. This occurs at $\nsi{ee} \approx \pm 1$ for $\varphi = \mp \pi/2$, essentially just beyond the reach of the LZ limit and suffers the same degradation in $\varphi$ as the other ER constraints.
Due to the additional charged-current contribution to \eves for electron neutrinos, the analytical condition for cancellation in ER is more involved than for NR. A detailed discussion of this can be found in Sec~IV.3 of Ref.~\cite{Amaral:2023tbs}. Nonetheless, as above, the loss of sensitivity in the ER channel can be compensated for by the complementary sensitivity in the NR channel.

\section{Conclusion}

We have presented Solar Neutrinos for Direct Detection (\snudd): an open-source Python package that enables the computation of the nuclear and electron recoil spectra at direct detection experiments. \snudd accounts for both solar and terrestrial neutrino propagation effects  as well as for modifications of the neutrino-nucleus and neutrino-electron scattering cross sections due to new physics.
In particular, the computation of the quantum-mechanical expectation value of the scattering cross sections via the trace formalism in \snudd ensures the correct treatment of neutrino flavour correlations and produces the correct neutrino scattering rates. \snudd allows the user to generate realistic direct detection signal predictions by incorporating detector efficiencies and energy resolution effects on the theoretical rates. Moreover, the modular structure of \snudd allows the user to implement custom cross sections, making it easily adaptable to incorporate new physics models beyond neutrino NSI.

We applied \snudd to compute new constraints in the NSI parameter space using both NR and ER data from the leading xenon-based direct detection experiments LZ, XENONnT, and PandaX-4T. We also derived the projected reach of a future xenon experiment based on the planned XLZD and PandaX-xT initiatives. We showed that current direct detection limits are rapidly approaching those obtained from global fit analyses using dedicated neutrino experiments. We found that the combination of NR and ER limits from direct detection experiment yields a powerful probe of NSI with the proton and electron. This can significantly alleviate blind spots plaguing single channel analyses.

Our work highlights that direct detection experiments are now in a position to provide complementary information to dedicated neutrino detectors. With \snudd, we provide the necessary tool to easily and consistently incorporate direct detection experiments into future NSI global analyses.


\acknowledgments

We would like to thank Peter Denton, Sk Jeesun, Michele Maltoni, Stephen Parke, Yuber P\'erez-Gonz\'alez, and Juehang Qin for helpful discussions.
We would also like to thank Teresa Marrod\'an and Giovanni Volta for help with understanding the XENONnT acceptance. 
DC, VC and PF would like to thank Valerie Domcke, Miguel Escudero, and the  CERN theory division  for their hospitality at CERN, where part of this work has been completed.

DA has been supported by ERC grant ERC-2024-SYG 101167211 by the European Union.
Views and opinions expressed are however those of the author(s) only and do not necessarily reflect those of the European Union, European Research Council Executive Agency, or other awarding body. Neither the European Union nor the granting authority can be held responsible for them. AC acknowledges the support of S. Ge, funded by the NSFC (Grant Nos. 12425506, 12375101, 12090060, and 12090064). The work of PF and VC was supported by a fellowship from La Caixa Foundation (ID 100010434 with code LCF/BQ/PR25/12110019). 
DC, PF, and VC acknowledge support from the Spanish Agencia Estatal de Investigaci\'on through the grants PID2024-155874NB-C22 (TheDeAs) and CEX2020-001007-S, funded by 
MCIN/AEI/10.13039/501100011033. 
This project has received funding from the European Union’s Horizon Europe research and innovation programme under the Marie Skłodowska-Curie Staff Exchange grant agreement No 101086085 – ASYMMETRY.
This work is partially funded by the European Commission – NextGenerationEU, through Momentum CSIC Programme: Develop Your Digital Talent. We acknowledge High Performance Computing support by Emilio Ambite, staff hired under the Generation D initiative, promoted by Red.es, an organisation attached to the Spanish Ministry for Digital Transformation and the Civil Service, for the attraction and retention of talent through grants and training contracts, financed by the Recovery, Transformation and Resilience Plan through the EU’s Next Generation funds.

\begin{appendices}

\crefalias{section}{appsec}
\crefalias{subsection}{appsec}

\appendix

\section{Details of solar neutrino propagation}
\label{app:solarnu}

In this section, we detail the calculation of the neutrino evolution through the solar medium outlined in~\cref{sec:solar_nu}. The main quantity we need to compute for this is the solar neutrino density matrix $\rho^\odot$, or more precisely the evolution $S$-matrix. To compute this we first have to find the diagonalisation of the effective Hamiltonian~\cref{eq:ham_eff} with the coefficients $\varepsilon_{N}^{\eta, \varphi}$ and $\varepsilon_{D}^{\eta, \varphi}$ being related to our parametrisation \cref{eq:nsi_def} by

\begin{equation}\label{eq:nsi_eff}
\begin{aligned}
\varepsilon_{D}^{\eta, \varphi} \equiv\,& c_{13}\, s_{13} \text{Re}\left(s_{23}\, \varepsilon_{e \mu}^{\eta, \varphi}+c_{23} \,\varepsilon_{e \tau}^{\eta, \varphi}\right)-\left(1+s_{13}^{2}\right) c_{23}\, s_{23}\, \text{Re} (\varepsilon_{\mu \tau}^{\eta, \varphi}) + \\
&-\frac{c_{13}^{2}}{2}\left(\varepsilon_{e e}^{\eta, \varphi}-\varepsilon_{\mu \mu}^{\eta, \varphi}\right)+\frac{s_{23}^{2}-s_{13}^{2}\, c_{23}^{2}}{2}\left(\varepsilon_{\tau \tau}^{\eta, \varphi}-\varepsilon_{\mu \mu}^{\eta, \varphi}\right)\,,  \\[.3cm]
\varepsilon_{N}^{\eta, \varphi} \equiv &c_{13}\left(c_{23} \,\varepsilon_{e \mu}^{\eta, \varphi}-s_{23}\, \varepsilon_{e \tau}^{\eta, \varphi}\right) \\
&+s_{13}\left[s_{23}^{2}\, \varepsilon_{\mu \tau}^{\eta, \varphi}-c_{23}^{2}\, (\varepsilon_{\mu \tau}^{\eta, \varphi})^*+c_{23}\, s_{23}\left(\varepsilon_{\tau \tau}^{\eta, \varphi}-\varepsilon_{\mu \mu}^{\eta, \varphi}\right)\right]\,.
\end{aligned}
\end{equation}

With the unitarity transformation defined in~\cref{eq:Urot} we can diagonalise the $2\times2$ Hamiltonian of~\cref{eq:ham_eff} via $H^{\mathrm{eff}}$ as $U^{m\dagger}_{12} H^{\mathrm{eff}} U^m_{12} = \operatorname{diag}(E^m_{1}, E^m_{2}) \equiv D(x)$. Defining the auxiliary quantities 
\begin{equation}\label{eq:pq}
\begin{aligned}
    & p =\sqrt{\left( \sin 2 \theta_{12} \cos \dCP + \frac{2 A_{cc}}{\Delta m^2} \xi(x) \text{Re}(\varepsilon_N)\right)^2+\left( \sin 2 \theta_{12} \sin \dCP + \frac{2 A_{cc}}{\Delta m^2}  \xi(x) \text{Im}(\varepsilon_N)\right)^2}\,,  \\
    & q = \cos 2 \theta_{12} + \frac{2 A_{cc}}{\Delta m^2} \xi(x) \varepsilon_D  - \frac{A_{cc}}{\Delta m^2} c_{13}^2 \,,
\end{aligned}
\end{equation}
where $A_{cc}=2\, E_\nu V(x)$,
we find the energy eigenvalues of the effective Hamiltonian in solar matter as
\begin{align}\label{eq:mat_energies}
    E^m_{1,2} =  \frac{A_{cc} c_{13}^2}{4 E_{\nu}} \pm \frac{\Delta m^2}{4 E_{\nu}} \sqrt{p^2 + q^2} \,.
\end{align}

The energy difference between these eigenvalues, responsible for the coherent mixing of the two matter mass eigenstates, is

\begin{equation}\label{eq:en_diff}
    \Delta E_{21}^m \equiv E^m_{2} - E^m_{1} = \frac{\Delta m^2}{2 E_{\nu}} \sqrt{p^2 + q^2}\,.\end{equation}  
Moreover, the matter mixing angle $\theta^{m}_{12}$  follows the relations

\begin{equation}\label{eq:mat_angles}
    \sin 2\theta_{12}^m = \frac{p}{\sqrt{p^2+q^2}}\,,  \qquad \cos 2\theta_{12}^m = \frac{q}{\sqrt{p^2+q^2}}\,,  \qquad  \tan 2\theta_{12}^m = \frac{p}{q}\,.
\end{equation}

With these expressions, we are finally in the position to describe the neutrino evolution in the full $3\times 3 $ picture. We can write the full neutrino evolution $S$-matrix through the Sun solving~\cref{eq:nu_evol} as 

\begin{equation}
    \tilde S
    = 
    \begin{pmatrix}
    \mathrm{Evol}[H^\mathrm{eff}] && 0 \\
    0 && \exp[-i \, \frac{\Delta m^2_{31}}{2\, E_\nu} L]
    \end{pmatrix}  \,.
\end{equation}

Using the slab approximation, we divide the path into $N$ infinitesimal slabs of thickness $\Delta x$. In this limit, the evolution operator is approximated by the ordered product of the individual evolution operators for each slab

\begin{equation}\label{evolution}
    \text{Evol}[H^{\mathrm{eff}}] =  \exp \left[ -i \int_{x_0}^{L} H^{\mathrm{eff}}(x) dx \right] \sim \lim_{\Delta x \to 0} \prod_{n=0}^{N} \exp \left[ -i H^{\mathrm{eff}}(x_n) \Delta x \right] \,.
\end{equation}

Diagonalising $H^{\mathrm{eff}}$ via the unitary transformation $U^m_{12}$ of~\cref{eq:Urot}, we find that when we combine the evolution across two consecutive slabs, we obtain the connection factor $U_n^{\dagger} U_{n+1} \approx \mathbb{1} + U_n^{\dagger} \dot{U}_{n} \Delta x$. Thus, the matter eigenstates, $\bm{\nu}_m$, change across two neighbouring slabs proportional to the combination  $U_n^{\dagger} \dot{U}_{n}$.
Hence, the $2\times2$ time-evolution matrix of the matter eigenstates is given by the diagonal part $D(x)=\operatorname{diag}(E^m_{1}, E^m_{2})$ and the non-adiabatic part $H_{\text{non-ad}}(x)=i\, U^{m\dagger}_{12}(x) \dot{U}^m_{12}(x)$,

\begin{equation}\label{fullssss}
\begin{aligned}
    i \frac{d}{dx} |\bm\nu_m\rangle & = \big[ \overbrace{D(x) + H_{\text{non-ad}}(x)}^{\tilde H(x)} \big] |\bm\nu_m\rangle \\ & =  \begin{pmatrix}
         E_1^m + \sin^2 \theta_{12}^m \ \dot \chi &  e^{i \chi}  [i \dot{\theta}_{12}^m - \sin \theta_{12}^m \cos \theta_{12}^m \ \dot \chi]\\ - e^{-i \chi} [i \dot{\theta}_{12}^m + \sin \theta_{12}^m \cos \theta_{12}^m \ \dot \chi]  & E_2^m - \sin^2 \theta_{12}^m \ \dot \chi
    \end{pmatrix}  |\bm\nu_m\rangle\,.
\end{aligned}
\end{equation}

In general, $\tilde H(x)$ will not be diagonal and hence has to be re-diagonalised at each slab across the path through the Sun. However, the solar matter profile obtained from simulations are smooth functions that vary slowly with radius. Thus, the off-diagonal elements of $\tilde{H}(x)$ are typically very small for solar propagation. If this is indeed the case, then $\tilde H $ is approximately diagonal along the entire path of the neutrinos through the Sun and the propagation is adiabatic.
To assess this more rigorously, we can define the adiabaticity parameter, $\gamma$, as the ratio of the diagonal to the off-diagonal entries of $\tilde H$,
\begin{equation}\label{eq:def_adiab}
     \gamma_{\pm}= \frac{|\frac{1}{2} \Delta E_{21}^m - \sin^2 \theta_{12}^m \ \dot \chi|}{|i \dot{\theta}_{12}^m \pm \sin \theta_{12}^m \cos \theta_{12}^m \ \dot \chi|}\,.
\end{equation} 
Then the adiabatic approximation is valid if the adiabatic parameter, $\gamma$, satisfies $\gamma_{\pm}\gg 1$.
The two different signs in the denominator correspond to the two different off-diagonal elements of the Hamiltonian.  We verified that the off-diagonal element with the positive sign is larger, and consequently computed $\gamma$ using this value.

In this approximation, 
the full $S$-matrix is given by
\begin{equation}\label{eq:sfull}
    S = 
    \begin{pmatrix}
    \exp\left[-i \,\int_0^L D(x) dx\right] && 0 \\
    0 && \exp[-i \,\Phi_{33}]
    \end{pmatrix} \,
    \underbrace{U^m_{12} (x_0)^\dagger O^\dagger}_{U^m_\mathrm{PMNS}(x_0)^\dagger}\,,
\end{equation}
where we evolve the neutrinos across the distance $L$ from their production point within the Sun, $x_0$,  to the solar surface. 
Given that in the adiabatic approximation the matrix $\tilde H(x)$ is thus approximately diagonal, and 
After a common re-phasing of the neutrino mass eigenstates, the upper $2\times2$ block in~\cref{eq:sfull} can be expressed as
\begin{equation}
     \exp\left[-i \,\smallint_0^L D(x) dx\right] \approx 
     \begin{pmatrix}
         e^{i\,\phi} & 0 \\
         0 & e^{- i\,\phi}
     \end{pmatrix} \,,
\end{equation}
which leads to the final expression for $S$ quoted in \cref{eq:Smatrixfull}.

\paragraph{Solar neutrino  density matrix (mass basis)}

From the evolution $S$-matrix, we can derive the expression for the three-flavour density matrix for solar neutrinos reaching Earth. With the projector onto the electron neutrino flavour, $\pi^{(e)}=\mathrm{diag}(1,0,0)$, the density matrix reads,
\begin{align}\label{eq:density_mat}
    \rho^{(e)} = S\, \pi^{(e)}\, S^\dagger = 
    \begin{pmatrix}
        |S_{11}|^2 &  S_{11}\, S_{21}^* & S_{11}\, S_{31}^* \\
        S_{11}^*\, S_{21} &  |S_{21}|^2 & S_{21}\, S_{31}^* \\
        S_{11}^*\, S_{31} &  S_{21}^*\, S_{31} & |S_{31}|^2 
    \end{pmatrix} \,.
\end{align}
With the expression for the solar evolution $S$-matrix in~\cref{eq:Smatrixfull}, we  can construct the density matrix according to~\cref{eq:density_mat} entirely from the three elements,
\begin{align}
    S_{11} & =    e^{i\,\phi}  \, c_{13} \,  c_{m}   \,,\\
    S_{21} & =  e^{-i\,(\phi+\chi)}  \, c_{13} \,  s_{m}\,,\\
    S_{31} & =  e^{-i\,\Phi_{33}}  \, s_{13}  \,,
\end{align}
where $c_{m}$ and $s_{m}$ refer to $\cos \theta_{12}^m$ and $\sin\theta_{12}^m$, respectively.
The six independent density matrix elements evolved to the surface of the Sun then read explicitly,
\begin{align}
    \rho_{11} =&~ c_{13}^2\ c_m^2\,, & \rho_{12} =&~ e^{i\,(2\phi+\chi)} \, c_{13}^2\ s_m\, c_m \,,  \\
    \rho_{22} =&~ c_{13}^2\ s_m^2 \,,  & \rho_{13} =&~ e^{i\,(\phi+\Phi_{33})} \, s_{13}\, c_{13}\  c_m \,,
    \\
    \rho_{33} =&~ s_{13}^2 \,, & \rho_{23} =&~ e^{-i\,(\phi-\Phi_{33}+\chi)} \, s_{13}\, c_{13}\  s_m  \,. 
\end{align}
As explained in the main text in~\cref{sec:solar_nu}, taking into  account the uncertainty in the neutrino production point effectively leads to an averaging of the phases $\phi$ and $\Phi_{33}$, resulting in $\langle e^{i  \phi} \rangle$ and $\langle e^{i  \Phi_{33}} \rangle \to 0$.  After further averaging the matter mixing angle, $\cos^2(\theta_{12}^m)$ over the solar neutrino production distribution, we recover the expression in \cref{eq:density_mass} for the mass-base neutrino density matrix. 

Expressing the density matrix in flavour space via the transformation in~\cref{eq:rho_solar}, we find the explicit expressions for the density matrix elements as,
\begin{equation}\label{eq:rho_flav}
\begin{aligned}
    \rho_{ee} =&~  s_{13}^4 +  c_{13}^4\, P_{\text{ee}}^{2 \nu }  \,, \\
    \rho_{\mu\mu} =&~ 
    c_{13}^2\, \Big[c_{23}^2\, \left(1-P_{\text{ee}}^{2 \nu }\right)+s_{13}^2 \,s_{23}^2 \,\left(1+P_{\text{ee}}^{2 \nu }\right) +\Delta_{\delta} \Big]  \,, 
    \\
    \rho_{\tau\tau} =&~ 
    c_{13}^2\, \Big[s_{23}^2\, \left(1-P_{\text{ee}}^{2 \nu }\right)+s_{13}^2 \,c_{23}^2 \,\left(1+P_{\text{ee}}^{2 \nu }\right) -\Delta_{\delta} \Big] \,, 
    \\
    \rho_{e\mu} =&~ c_{13}\, s_{13}^3\, s_{23}-\frac{1}{2}\, c_{13}^3\, \bigg[2\, s_{13}\, s_{23} \, P_{\text{ee}}^{2 \nu } +  c_{23}
    \sin 2 \theta _{12} \cos 2 \theta^m_{12} e^{i \dCP} \bigg] \,, 
    \\
    \rho_{e\tau} =&~ c_{13}\, s_{13}^3 \, c_{23} - \frac{1}{2}\, c_{13}^3\, \bigg[2\, s_{13}\, c_{23} \, P_{\text{ee}}^{2 \nu }-   s_{23}\,  \sin 2 \theta _{12} \cos 2 \theta^m_{12} e^{i \dCP} \bigg] \,, 
    \\
    \rho_{\mu\tau} =&~ \frac{1}{2}\, c_{13}^2 \Big[ \sin \left(2 \theta _{23}\right)  \Big(\left(1+s_{13}^2\right)\,P_{\text{ee}}^{2 \nu } -c_{13}^2 \Big) + 2\,  \cot 2 \theta _{23} \Delta_{\delta} + \\ \quad &~  -i \sin (\dCP) s_{13} \sin 2 \theta _{12} \cos 2 \theta^m_{12} \Big] \,,
\end{aligned}
\end{equation}

where we have defined 

\begin{equation}
    \begin{aligned}
            P^{2\nu}_{ee} &= \frac{1}{2}\, \Big(1+\cos 2 \theta _{12} \,\cos 2 \theta^m_{12}\Big)=c_{12}^2 c_m^2 + s_{12}^2 s_m^2 \,, \\
    \Delta_{\delta} &= \frac{1}{2}\  s_{13} \, \sin 2 \theta _{12}\, \sin 2 \theta _{23}\, \cos 2 \theta^m_{12}\cos\dCP\,. 
    \end{aligned}
\end{equation}

\subsection{Adiabaticity of neutrino evolution in the Sun with NSI}\label{app:adiab}

When computing the neutrino density matrix for solar propagation, we assume adiabatic evolution of the neutrino states in the Sun. 
In order to asses the validity of this assumption, we have  carefully re-derived the adiabaticity parameter $\gamma$ in~\cref{eq:def_adiab}, including the effect of complex-valued NSI parameters $\varepsilon_{\alpha\beta}$ and a non-zero CP-phase, $\dCP$.
The adiabatic approximation is valid as long as the adiabaticity parameter, $\gamma$, satisfies, $\gamma\gg1$. With the help of~\cref{eq:def_adiab}, we can asses this condition both when varying the NSI parameters and the CP-phase.

\begin{figure}[t]
\centering
\includegraphics[width=\textwidth]{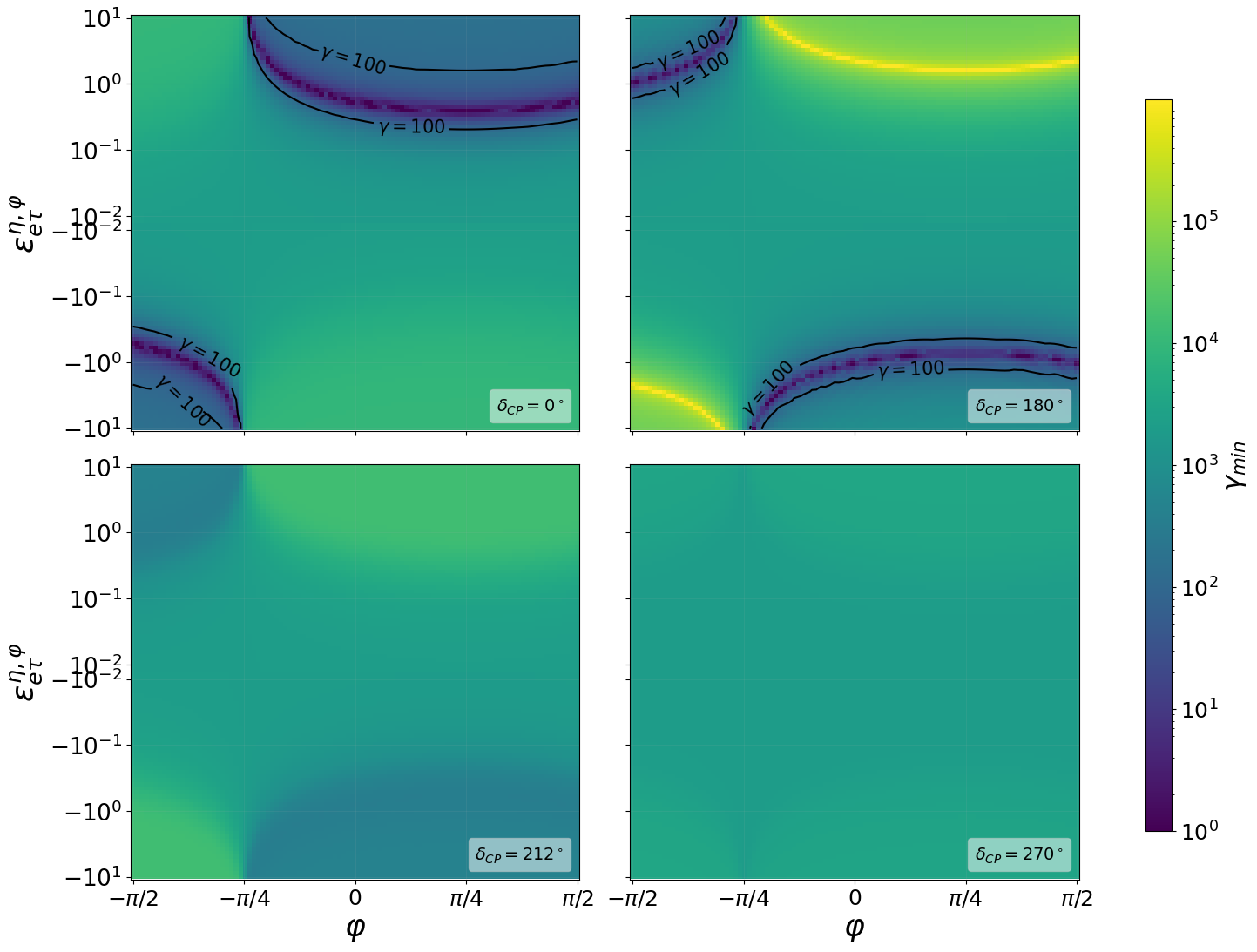}
\caption{Minimum adiabaticity parameter $\gamma_{\text{min}}$ for solar neutrinos ($^8\text{B}$ flux) as a function of the angle $\varphi$ and the off-diagonal NSI parameter $\nsi{e\tau}$. The calculation is performed at a fixed neutrino energy $E_{\nu} = 10$ MeV, assuming $\eta = 0$. The colour scale represents the magnitude of $\gamma_{\text{min}}$, where lower values (darker regions) indicate a departure from the adiabatic regime. The black contour highlights the critical threshold $\gamma = 100$, identifying the regions of the parameter space where the adiabatic approximation begins to fail. The y-axis is shown in symmetric logarithmic scale to highlight the sensitivity across different orders of magnitude for $\nsi{e\tau}$. We show four plots with different $\dCP$ values: $\dCP = 0^\circ$ (CP-conserving case), $\dCP = 180^\circ$ (CP-conserving flip), $\dCP = 212^\circ$ (best-fit asymmetry), $\dCP = 270^\circ$ (maximum CP violation).}
\label{fig:gam_phi}
\end{figure}

In~\cref{fig:gam_phi}, we show the minimum values of the adiabaticity parameter $\gamma_{\rm min}$ across the solar profile as a function of the NSI magnitude $\varepsilon_{e\tau}$ and the angle in the charged plane, $\varphi$, for four different values of $\dCP$. We illustrate the general behaviour of $\gamma$ for the example of $(\varepsilon_{e\tau}, \varphi)$, but we have checked that the conclusion we draw here are generic and apply also to the other combinations $(\varepsilon_{\alpha\beta}, \eta)$ and $(\varepsilon_{\alpha\beta}, \varphi)$.
Note also that the adiabaticity parameter is decreasing with energy, since $\gamma \propto \Delta m^2 / 2 E_{\nu}$ via~\cref{eq:en_diff}. The presence of the derivative $\dot \chi$ for $\dCP \neq 0$ dampens this energy dependence, but the general trend persists. Hence, the minimum values are attained at the highest neutrino energies (i.e.~$^8B$ and $hep$ neutrinos).

Looking at the top panels of~\cref{fig:gam_phi}, we can see that in the case of CP-conserving $\dCP$,  there exists a band of $\varepsilon_{e\tau}$ values for which $\gamma\leq100$ (solid black lines), which we take as a conservative threshold for adiabaticity, and ultimately even approaches $\gamma\sim 1$ (dark blue).
We can further observe that $\gamma$  seems to obey a slightly broken reflection symmetry for shifts of $\pi$ in $\dCP$, i.e.~it is symmetric under the joint transformation
\begin{equation}\label{eq:tr}
    \varepsilon_{e\tau} \to -\varepsilon_{e\tau}\,, \qquad  \quad \dCP \to \dCP + \pi\,.
\end{equation}
However, this reflection symmetry is not exact. To understand this we will have a closer look at $\gamma$, which 
is given by the ratio of the energy splitting of the matter eigenvalues, $\Delta E_{21}^m $, and the derivative of the matter mixing angle, $\dot \theta_{12}^m$, both terms $\pm$ a term depending on the derivative $\dot \chi$. Since $\chi$ is symmetric under the transformation~\cref{eq:tr}, we will focus only on $\Delta E_{21}^m = 2 \Delta \sqrt{p^2 + q^2}$ and $\dot \theta_{12}^m$. Both these quantities depend on the parameters $p$ and $q$ defined in \cref{eq:pq}. Hence, the breaking of the reflection symmetry must originate in the symmetry properties of $p$ and $q$.
While all terms in $p$ depend linearly both on $\varepsilon_{\alpha\beta}$ and $\dCP$, we have $p (-\varepsilon_{e\tau}, \dCP + \pi) = p (\varepsilon_{e\tau}, \dCP)$. However, $q$ contains a term that only depends on $\varepsilon$, and hence spoils  the reflection symmetry.

The sharp maximum in the adiabaticity parameter (yellow bands) observed in the upper right panel of~\cref{fig:gam_phi} is due to a near cancellation in the denominator of $\gamma$. Analysing \cref{eq:def_adiab}, we find that the denominator is dominated by the derivative of the mixing angle, $\dot{\theta}_{12}^m$ (see~\cref{eq:mat_angles}), while the contribution from $\dot{\chi}$ is negligible. Consequently, the resonance behaviour is entirely driven by $\dot{q} \propto [f'(x)\varepsilon_D - \frac{1}{2} c_{13}^2 V'(x)]$, where $f'(x)=V'(x) \xi (x) + V(x) \xi'(x)$. The maximum in $\gamma$ occurs precisely for values of $\varepsilon_D$ for which these two terms in $\dot q$ cancel.

Finally, we see in the bottom panels of~\cref{fig:gam_phi} that for the best fit value, $\dCP=212^\circ$, and for maximum CP-violation, $\dCP=270^\circ$, the adiabaticity parameter is safely above $\gamma>100$ across the entire parameter space, making the adiabatic approximation safe to employ across the entire parameter space.

\section{Details of Earth propagation}

\subsection{Earth's motion: two-body problem redux}
\label{app:Earth_motion}

In order to incorporate earth matter effects on the solar neutrino signal in terrestrial detector experiments, we need to track
the time dependent path of the neutrinos through the Earth. If we assume a spherically symmetric earth matter profile, this path is completely determined by the incident angle of neutrinos under which they appear at the detector. This is governed by Earth's motion around the Sun and Earth's rotation.

To track Earth's relative motion around the Sun, we have to solve the two-body problem described by the system of coupled first-order differential equations,
\begin{align}\label{eq:ode1}
    \dot{\mathbf{r}}=
    \begin{pmatrix}
    \dot x \\
    \dot y
    \end{pmatrix} &= 
    \begin{pmatrix}
    u \\
    v
    \end{pmatrix} \,, \\[4pt]
    \ddot{\mathbf{r}}=
    \begin{pmatrix}
    \dot u \\
    \dot v
    \end{pmatrix}  &=
    -\frac{\mu}{r^3}
    \begin{pmatrix}
    x \\
    y
    \end{pmatrix} = -\frac{\mu}{r^3} \mathbf{r}\,, \label{eq:ode2}
\end{align}
where $\mu= G M$. 

In order to solve these differential equations, we need to specify the initial conditions. At the perihelion, the Earth's position is given by $\mathbf{r}=(a(1-e),0)$, where $a=1$ AU is the length of the semimajor axis of Earth's orbit and $e=0.0167$ is its eccentricity.  We can find the initial velocity from the (constant) orbital energy $\varepsilon=v^2/2-\mu/r$, which for a closed orbit is equal to $\varepsilon=-\mu/(2a)$. Thus, the initial velocity at the perihelion is given by
\begin{equation}
    \dot{\mathbf{r}}=
    \begin{pmatrix}
        0 \\
        \sqrt{\mu\left(\frac{2}{r}-\frac{1}{a}\right)}\,.
    \end{pmatrix}
\end{equation}
With this we can solve the coupled system of equations~\cref{eq:ode1,eq:ode2} and obtain Earth's trajectory $\mathbf{r}=(x(t),y(t))$. From this we can obtain the true anomaly $\alpha$ (the angle of the Earth around the Sun taken w.r.t.~the perihelion) as
\begin{equation}
    \alpha(t) = \tan^{-1} \left(\frac{y(t)}{x(t)}\right)\,.
\end{equation}
This means that in the solar reference frame with the Sun at the origin, the incident direction of the solar neutrinos is given by the unit vector 
\begin{equation}
    \bm{\hat{\nu}}=
    \begin{pmatrix}
        \cos \alpha(t) \\
        \sin \alpha(t) \\
        0
    \end{pmatrix}\,.
\end{equation}

Now want to see how this translates into the time-varying incident angle (nadir) $\eta_{\rm nad}$ of neutrinos at a given detector location on Earth. Let us consider a detector at a latitude $\theta_{\rm det}$. Then its time-varying position on Earth's surface in the Earth reference frame (rotation axis taken to be $z$-axis) is given by
\begin{align}
    \boldsymbol{P_\oplus}(t)= R_\oplus
    \begin{pmatrix}
        \cos \phi(t)\, \sin \theta_{\rm det} \\
        \sin \phi(t)\, \sin \theta_{\rm det} \\
        \cos \theta_{\rm det}
    \end{pmatrix} &&, \ \text{with} \ \phi(t) =  \frac{2 \pi t}{T_\oplus}\,,
\end{align}
and $R_\oplus$ and $T_\oplus$ are Earth's radius and rotation period, respectively.  We can rotate into the solar plane about the ecliptic angle $\theta_\oplus$,
\begin{align}
    \boldsymbol{P_\odot}(t)= 
    \begin{pmatrix}
    \cos \theta_\oplus & 0 & \sin \theta_\oplus \\
    0 & 1 & 0\\
    - \sin \theta_\oplus & 0 & \cos \theta_\oplus
    \end{pmatrix}
    \boldsymbol{P_\oplus}\,.
\end{align}

The incident nadir angle $\eta_{\rm nad}$ is determined by the scalar product of the neutrino incident direction, 
$\hat{\boldsymbol{\nu}}$ 
and the unit normal vector at the detector location, 
$\hat{\boldsymbol{n}}=\boldsymbol{P_\odot}/R_\oplus$,
\begin{equation}
	\cos \left(\eta_{\rm nad}(t)\right) = \hat{\boldsymbol{\nu}} \cdot \hat{\boldsymbol{n}}
	= c_{\alpha(t)} (c_{\theta_\oplus}\, s_{\theta_{\rm det}}\, c_{\phi(t)} + s_{\theta_\oplus}\,c_{\theta_{\rm det}}) + s_{\theta_\oplus}\, s_{\theta_{\rm det}}\, s_{\phi(t)} \,,
\end{equation}
with $s_x$ and $c_x$ denoting the $\cos$ and $\sin$, respectively.

\begin{figure}[t]
\centering
\begin{tikzpicture}[scale=1, line cap=round, line join=round]
  \def\Re{3.0}          
  \def\Rt{3.25}         
  \def\etadeg{30}        
  \def\xv{-1.40}        

  \pgfmathsetmacro{\ycap}{sqrt(\Re*\Re - \xv*\xv)}         
  \coordinate (O) at (0,0);
  \coordinate (I) at (\xv,  {sqrt(\Rt*\Rt-\xv*\xv)});                        
  \coordinate (F) at (\xv, -\ycap);                        
  \coordinate (M) at (\xv, 0);                             
  \pgfmathsetmacro{\pf}{0.25}
  \coordinate (P) at ($(F)!\pf!(I)$);

  \pgfmathsetmacro{\dx}{sin(\etadeg)}
  \pgfmathsetmacro{\dy}{cos(\etadeg)}
  \coordinate (Qfar) at ($(F)!6!(F)+(\dx,\dy)$);           

  \path[name path=earth]  (O) circle (\Re);
  \path[name path=atm]    (O) circle (\Rt);
  \path[name path=traj]   (F) -- (Qfar);


  \draw[very thin] (O) circle (\Rt);
  \draw[line width=0.9pt] (O) circle (\Re);

  \draw[line width=0.9pt] (I) -- (F);


  \draw[thin,gray!60] (O) circle (0.65*\Re);
  \draw[thin,gray!60] (O) circle (0.90*\Re);

  \fill (O) circle (1.2pt) node[below right=-2pt] {$O$};
  \fill (I) circle (1.6pt) node[left=3pt] {$I$};
  \fill (F) circle (1.6pt) node[right=6pt] {$F$};
  \fill (M) circle (1.6pt) node[left=3pt] {$M$};
  \fill (P) circle (1.6pt) node[left=3pt] {$P$};

  \draw[dashed] (O) -- ($(M)!1!(O)$ -| \xv,0) node[pos=0.55, above] {};
  \draw[dashed] (O) -- (M) node[midway, above] {};

  \draw[dashed] (O) -- (P) node[midway, above left=-2pt] {$r$};

  \draw[dashed] (F) -- ($ (O)!-1!(F) $);

  \pic[draw,->,angle radius=120pt,angle eccentricity=1.07,
     "\text{\ensuremath{\eta_\mrm{nad}}}"] {angle = O--F--I};

  \path let \p1 = ($(F)-(I)$) in
    coordinate (Iv) at ($(I)+(0,-0.8)$); 


  \node[align=left, anchor=west] at ($(O)+(4.1,1.1)$) {%
    $I$ = $\nu$ entry point\\
    $F$ = $\nu$ endpoint (detector)\\
    $M$ = midpoint of trajectory\\
    $P$ = generic $\nu$ position\\
    $x=IP$ (trajectory coordinate)\\
    $r=OP$ (radial distance)\\
    $\eta_\mrm{nad}$ = nadir angle};

\end{tikzpicture}
\caption{Diagram showing the neutrino path through the Earth. Note that the number of shells is illustrative and that the PREM model~\cite{Dziewonski:1981xy} contains 11 shells. }
\label{fig:EARTH}
\end{figure}

\subsection{Neutrino trajectories through Earth}
\label{app:nutrajEarth}

Here, we describe the details of our determination of a neutrino's path through the Earth. We adopt the geometry of Ref.~\cite{Lisi:1997yc}, whereby the nadir angle $\eta_\mrm{nad}$ is the angle between the upward direction from the detector and the incident neutrino (see~\cref{fig:EARTH}). A neutrino that travels through the centre of the Earth to the detector has $\eta_\mrm{nad}=0$. Therefore, the neutrino travels through some portion of the Earth when $-\pi/2<\eta_\mrm{nad}<\pi/2$. 

To calculate the trajectory of a neutrino, we take the detector coordinate to be $\mathbf{F}=(-R_\oplus\sin\eta_\mrm{nad}, -R_\oplus\cos\eta_\mrm{nad})$, where $R_\oplus$ is the Earth radius. The neutrino direction is taken as $\hat{\mathbf{n}}=(0,-1)$. Then the backward ray from the detector can be parametrised via the parameter $t$, $\mathbf{x}(t)=\mathbf{F}-t\,\hat{\mathbf{n}}$. Matching this path from the detector to the top of the atmosphere, $\vert\mathbf{x}(t)\vert^2=R_{\rm tot}^2 = (R_\oplus+R_{\rm ATM})^2$ gives
\begin{equation}
    t_1 = R_\oplus\cos\eta + \sqrt{R_{\rm tot}^2+R_{\oplus}^2\left(\cos^2\eta_\mrm{nad}-1\right)}\,.
\end{equation}
Along the trajectory, we slice $N_{\rm s}$ times and evolve the Hamiltonian given in~\cref{eq:hamiltonian_nsi} as described in~\cref{sec:earth_matter}. For each slice, we determine the electron density using PREM~\cite{Dziewonski:1981xy}, which is defined in radial coordinates with the origin at the centre of the Earth. To relate the neutrino's entry point in the atmosphere to its arrival at the detector, we account for the difference between the entry angle $\theta$ and the nadir angle. This is achieved by conserving the impact parameter (denoted as OM in~\cref{fig:EARTH})
\begin{equation}
    R_\oplus\sin\eta_\mrm{nad}=R_{\rm tot}\sin{\theta},
\end{equation}
which leads to
\begin{equation}
    \cos\theta =\sqrt{1-\left(\frac{R_\oplus}{R_{\rm tot}}\right)^2\sin^2\eta_\mrm{nad}}\,.
\end{equation}
Then using the law of cosines, we can determine the radial distance of a point along the trajectory, $x\in[0,1]$,
\begin{equation}
    r(x)^2=R_{\rm tot}^2 + x^2R_{\rm tot}^2 - 2 R_{\rm tot}^2 x \cos\theta \, .
\end{equation}
With this, we can calculate the matter Hamiltonian~\cref{eq:hamiltonian_nsi} at any point along the neutrino's path through the Earth, which is required for solving the neutrino propagation~\cref{sec:earth_matter}.

\end{appendices}

\bibliographystyle{JHEP}
\bibliography{biblio}

\end{document}